\documentclass[a4paper]{aa}
\usepackage{txfonts,graphicx}
\begin{document}

\title{Hydrogen and helium line formation in OB dwarfs and giants
\thanks{Based on observations collected at the European Southern Obser\-vatory,
Chile (ESO N$^{\circ}$ 074.B-0455(A)) and at
the Centro Astro\-n\'omico Hispano Alem\'an at Calar Alto, operated jointly
by the Max-Planck Institut f\"ur Astronomie and the Instituto de Astro\-f\'isica
de Andaluc\'ia.}\fnmsep\thanks{Table~\ref{tablelines}
and Figs.~\ref{visualHR1861}--\ref{visualHR5285} are only available in electronic form at
{\tt http://www.edpsciences.org}}}
\subtitle{A hybrid non-LTE approach}
\author{M. F. Nieva\inst{1,2} \and N. Przybilla\inst{1}}

\offprints{M.F. Nieva~(nieva@sternwarte.uni-erlangen.de)}
\institute{Dr. Remeis Sternwarte Bamberg, Sternwartstr. 7, D-96049 Bamberg, Germany
\and Observat\'orio Nacional, Rua General Jos\'e Cristino 77
CEP 20921-400, Rio de Janeiro, Brazil\\}
\date{Received{\ldots}; accepted{\ldots}}
\abstract{}{
Hydrogen and helium line spectra are crucial diagnostic features for the
quantitative analysis of OB stars. Hybrid non-LTE line-formation calculations for 
these elements have not been discussed thoroughly so far,
despite their wide use for analyses of metal line spectra.
We compute synthetic spectra based on a hybrid 
non-LTE approach in order to test the ability of these models to reproduce high-resolution 
and high-S/N spectra of dwarf and giant stars and also to compare them with
published grids of non-LTE (OSTAR2002) and LTE (Padova) models. 
}
{
Our approach solves the restricted non-LTE problem based on classical
line-blanketed LTE model atmospheres. State-of-the-art model atoms and
line-broadening theories are employed to model the H and
\ion{He}{i/ii} spectra over the entire optical~range~and~in~the~near-IR.
}
{
A comparison with published line-blanketed non-LTE models
validates the suitability of the LTE approximation for modelling the 
atmospheric structure of late O to early B-type dwarf and giant stars at metallicities down to (at least)
1/5\,$\times$\,solar. Our hybrid non-LTE synthetic spectra simultaneously match 
almost all measurable hydrogen and helium lines
observed in six test stars over a wide spectral range from the Balmer limit
to the near-IR, except for only a few well-understood cases. A robust starting
point for studies of the metal spectra is thus established.
Our approach reproduces other published non-LTE calculations, but avoids 
inconsistencies in the modelling of the \ion{He}{i} singlets. These have
recently been discussed in the literature in the context of O-type stars and 
we show that they persist in the early B-types. 
Our approach improves on published pure LTE models -- widely applied for OB star analyses
-- in many aspects: 
non-LTE strengthening and the use of improved line-broadening data result in 
 significant differences in the line profiles and equivalent widths of 
the Balmer and helium lines.
Where possible, systematic effects on the stellar parameter determination are
 quantified, e.g. gravities derived from the H$\gamma$ wings may be 
overestimated by up to $\sim$0.2\,dex for dwarfs at our upper temperature boundary 
of 35\,000\,K in LTE.}
{}
\authorrunning{Nieva \& Przybilla}
\titlerunning{H and He line formation in OB dwarfs and giants}
\maketitle

\section {Introduction}
Massive stars are the main drivers for the dynamical and chemical evolution of
the interstellar medium (ISM), and thus for the evolution of entire galaxies.
They are important sites of nucleo\-synthesis and main contributors to the
energy and momentum budget of the ISM, as sources of ionizing UV radiation,
through stellar winds and their final explosion as supernovae.

Quantitative analyses of O- and B-type stars can provide observational constraints
on both stellar and galactochemical evolution. In the first case,
information on basic stellar parameters and mixing with nuclear-processed
matter (abundances of the light elements) facilitates empirical
evaluation of the quality of different evolutionary models (see e.g.
Maeder \& Meynet~\cite{MaMe00}). In the second
case, information on the spatial distribution of present-day elemental abundances
allows Galactic metallicity gradients to be derived (e.g. Gummersbach et
al.~\cite{Gummersbachetal98}; Rolleston et al.~\cite{Rollestonetal00};
Daflon \& Cunha~\cite{DaCu04}) as the basis for galactochemical evolution
modelling (e.g. Hou et al.~\cite{Houetal00}; Chiappini et
al.~\cite{Chiappinietal01}; 
Cescutti et al.~\cite{Cescuttietal06}). Stellar data can hereby complement and
independently verify analogous abundance studies of \ion{H}{ii} regions (e.g. Shaver
et al.~\cite{Shaveretal83}). Using the present generation of large
telescopes and high-resolution spectrographs, pristine abundances can be derived
not only for Galactic early-type stars, but also for objects
in the metal-poor environments of the Magellanic Clouds (e.g. Korn et al.~\cite{Kornetal02},
\cite{Kornetal05}; Rolleston et al.~\cite{Rollestonetal03}; Hunter et
al.~\cite{Hunteretal06}).

In order to be meaningful, the parameter and abundance determination for
individual objects has to be unbiased by systematic error.
Early and mid O-type stars pose considerable challenges
to model atmosphere analyses, because of sphericity, mass-outflow, and
non-LTE line-blocking/blanketing effects.
For these reasons the least luminous (but most numerous) massive stars,
late O and early B-type (OB) stars (Jaschek \& Jaschek~\cite{jj90}), 
have been at the focus of abundance studies for a long
time (e.g. Gies \& Lambert~\cite{GiLa92}; Kilian~\cite{Kilian92}; Cunha \&
Lambert~\cite{CuLa92}; and numerous similar studies ever since).
Overall, the atmospheres of OB dwarf and giant stars are supposed to be described reasonably well by
one-dimensional, plane-parallel, homogeneous, and hydrostatic line-blanketed
LTE models in radiative equilibrium. However, this does not imply that
quantitative analyses of the spectra of OB dwarfs and giants are trivial.

Most of the published quantitative analyses follow two approaches.
Either they are based on pure LTE analyses (e.g. Rolleston et al.~(\cite{Rollestonetal00}), 
and references therein) or they solve the restricted non-LTE problem on prescribed line-blanketed LTE
model atmospheres. Typically, metal ions are considered in this case. A simultaneous 
treatment of H and He is exceptional (e.g. Gummersbach et
al.~\cite{Gummersbachetal98}; Morel et al.~\cite{Moreletal06}). 
A few studies use unblanketed non-LTE atmospheres with subsequent non-LTE 
line formation (e.g. Rolleston et al.~\cite{Rollestonetal00}), and only
recently have grids of fully line-blanketed non-LTE models become available
(Lanz \& Hubeny~\cite{la-hu03}: concentrating on O-type stars but
extending into the B star regime; 
Dufton et al.~\cite{Duftonetal05}: concentrating on B-type stars, not
publically available). A thorough test of the models regarding their ability to 
reproduce the H and He spectra via direct comparison with observation
throughout the entire visual range and in the near-IR 
in OB dwarfs and giants covering a broad parameter range has not been
published so far (note their availability -- at least for subsets of
lines -- for O stars: Bouret et al.~\cite{Bouretetal03}; 
Repolust et al.~\cite{Repolustetal04}, \cite{Repolustetal05}; 
Mokiem et al.~\cite{Mokiemetal05}, \cite{Mokiemetal06}; 
Heap et al.~\cite{Heapetal06}). Having this kind of test available would help the user of grids 
to understand the strengths and the limitations of the models. Usually, only one or two hydrogen 
Balmer lines and selected helium lines in the optical blue are considered
in the literature.

Hydrogen and helium are of major interest in the astrophysical context, as
they constitute practically all light-emitting plasma. The lines of hydrogen
and helium are the strongest spectral features in OB stars. Inasmuch as they are
primary diagnostic tools for stellar analyses throughout the
Hertzsprung-Russell diagram, they sample the plasma conditions
throughout large parts of the stellar atmosphere, to a far greater extent than
do the metal lines. However, the information
on the temperature and density structure encoded in the spectra has to be
{\em interpreted}. This is done by comparison with synthetic spectra, which
requires that the basic atmospheric structure equations in combination with the radiative
transfer problem be solved. The model predictions may differ,
depending on the approximations made and on the atomic data used. Their
quality can be assessed by their ability to
reproduce observation. In the optimum case {\em all} observational constraints
(continua/spectral energy distribution, line profiles) should be reproduced {\em
simultaneously}, indicating the absence of systematic error (assuming
a unique solution). A thorough reproduction of the hydrogen and
helium line spectra should therefore be viewed as a precondition for all further studies.

Non-LTE effects play a dominant r\^ole in the
formation of the hydrogen and helium line spectra in early-type stars, as
known since the seminal work by Auer \& Mihalas~(\cite{AuMi72}, \cite{AuMi73}).
Despite the enormous progress made over the past thirty years, some notorious problems have
remained. Observations in the (near-)infrared provide one key to improving
the situation via extension of our observational database to a domain of
amplified non-LTE effects (in OB stars). Some of the problems have been
related recently to the remaining inaccuracies in the atomic data.
 Thus, the modelling of the hydrogen Paschen, Brackett, and Pfund
series in early-type stars could be improved, resulting in corrections of
equivalent widths by as much as a factor 2--3 (Przybilla \&
Butler~\cite{przyb-but04}; but see also Repolust et al.~\cite{Repolustetal05}).
Also, the observed behaviour of the
\ion{He}{i}~$\lambda$10\,830\,{\AA} transition in OB dwarfs could be
reproduced for the first time (Przybilla~\cite{przyb05}).
In other cases, the reasons for shortcomings in the non-LTE modelling can be
subtler. An example of this is the \ion{He}{i} singlet line problem
in early-type stars: computations with non-LTE model-atmosphere codes
reveal discrepancies not only between theory and observation but also
between different theoretical calculations.
The overlap of an \ion{He}{i} resonance transition with \ion{Fe}{iv}
lines results in high sensitivity to the model assumptions (Najarro et
al. \cite{Najarroetal06}).

The aim of this paper is to evaluate the status of non-LTE line-formation
computations for the two most abundant elements in the most common targets of
massive star analyses, OB dwarf and giant stars. This work 
constitutes the basis for further studies of the metal line spectra 
(e.g. Nieva \& Przybilla~\cite{np06b},~\cite{np07}).
First, we test our hybrid non-LTE approach (Sect.~2) thoroughly 
on high-quality spectra of six stars in the solar vicinity (Sects.~3 \&~4).
In contrast to typical studies from the literature, we investigate
the {\em entire} hydrogen and helium line spectra in the optical range,
plus some additional near-IR data, taking advantage of our recently
improved non-LTE model atoms. After making sure that excellent agreement
between theory and observation can be obtained (i.e. also avoiding the
aforementioned \ion{He}{i} singlet line problem), we compare our modelling
with libraries of synthetic fluxes from the literature (Sect.~5). This is done
 in order to test
their suitability for quantitative analyses of OB dwarfs and giants.
Such libraries are required for (automatised) analyses of large
observational datasets obtained with existing or future
multi-object spectrographs (e.g. the VLT-FLAMES survey of massive stars:
Evans et al.~\cite{Evansetal05}; GAIA: Perryman et al.~\cite{Perrymanetal01}).
It is shown that reliable modelling of the line spectra of the two most
basic elements is not straightforward. On the contrary, considerable systematic
errors may result for quantitative analyses of OB stars when applying
these libraries blindly. Finally, the results of our investigation are
discussed (Sect.~6) and the main conclusions summarised (Sect.~7).

\section{Model calculations}
The hybrid non-LTE approach solves the restricted non-LTE problem on the
basis of prescribed LTE model atmospheres. The approach is physically less elaborate
than fully self-consistent non-LTE calculations, as more approximations
are involved. However, at the same time it is superior to the pure LTE
approximation. In particular, it provides an {\em efficient} way to compute
realistic synthetic spectra in all cases where the atmospheric structure is close
to LTE (which puts restrictions on the parameter space coverage). 
The hybrid non-LTE approach also allows extensive non-LTE model 
atoms to be implemented, facilitating a highly detailed treatment of the 
atomic processes involved (e.g. account for the resolved resonance structure
in photoionizations, avoidance of the -- powerful, however also approximate -- 
superlevel formalism).

We compute line-blanketed, plane-parallel,
homogeneous, and hydrostatic LTE model atmospheres using the ATLAS9 code
(Kurucz~\cite{kurucz93b}). Non-LTE population numbers and synthetic spectra are then 
obtained with
recent versions of DETAIL and SURFACE (Giddings~\cite{gid81}; Butler \&
Giddings~\cite{but_gid85}). The coupled~radiative transfer and statistical
equilibrium equations are~solved with DETAIL, employing the Accelerated Lambda Iteration 
(ALI) scheme of Rybicki \&
Hum\-mer~(\cite{rh91}). Synthetic spectra are calculated with SURFACE,
using refined line-broadening theories.

The non-LTE model atoms for hydrogen and \ion{He}{i/ii} adopted in the present
work are described in detail by Przybilla \& Butler (\cite{przyb-but04}) and
Przybilla~(\cite{przyb05}), respectively. Use of improved atomic data
for electron impact excitations, in particular from {\em ab-initio}
computations, allows consistent results from the hydrogen lines in
the visual and near-IR to be derived throughout the entire range of early-A to O stars .
A 15-level model is used for modelling
main sequence stars, as well as a 20-level model for the giants.
The \ion{He}{i/ii} model atom considers all \ion{He}{i} $LS$-coupled terms
up to the principal quantum number $n=5$ individually, and packed levels up to
$n=8$ separately for the singlet and triplet spin system. All levels up to
$n=20$ are considered in the \ion{He}{ii} model. This model atom has been
successfully used to reproduce observed trends of the highly non-LTE-sensitive
\ion{He}{i}\,$\lambda$10\,830\,{\AA} transition in early-type main sequence
stars (Przybilla~\cite{przyb05}). The \ion{He}{i/ii} model was also employed to analyse the visual/near-IR spectra of extreme helium stars
(Przybilla et al.~\cite{Przybillaetal05}) and subluminous B stars (Przybilla
et al.~\cite{Przybillaetal06a}).

\begin{figure}[h]
\centering
\includegraphics[width=.98\linewidth,height=5.3cm]{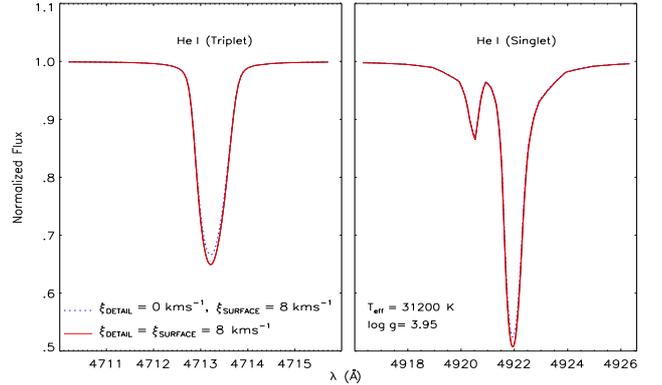}\\
\caption{Sensitivity of theoretical \ion{He}{i} line profiles to modifications of the
microturbulent velocity in the statistical equilibrium calculations.
The microturbulence of the ATLAS9 model atmosphere structure is held fixed at
$\xi$\,$=$\,8\,km\,s$^{-1}$. The test calculations have been done for one of our sample
stars (HR\,3055). Similar effects are found for other \ion{He}{i} lines,
while
the \ion{He}{ii} and H lines are practically insensitive to this.}
\label{det_mic}
\end{figure}

Radiative transitions in DETAIL are treated with simplified line broadening
formalisms: for transitions between hydrogen levels with $n\le7$ approximate 
Stark-broadening (Griem~\cite{Griem60}, following the implementation of Auer \&
Mihalas~\cite{AuMi72}, Appendix) is considered, while for all other
transitions, also in \ion{He}{i/ii}, depth-dependent Doppler profiles are
assumed. Microturbulence is explicitly accounted for by including
the appropriate term in the Doppler width
\begin{equation}
\Delta \lambda_\mathrm{D} =
\frac{\lambda_\mathrm{0}}{c}\,\left( \varv^2_\mathrm{th} + \xi^2
\right)^{1/2}~,
\end{equation}
where $\lambda_0$ is the rest wavelength of the transition, $c$ the speed of light,
$\varv_\mathrm{th}$ the thermal velocity for the chemical species of
interest and $\xi$ the microturbulent velocity. Both continuous absorption
and line blocking (via Kurucz' opacity distribution functions, ODFs,
Kurucz~\cite{kurucz93a}, using the `little' wavelength interval versions)
are accounted for as background opacities in solving the radiation transfer.
 In this regard the hybrid non-LTE approach has an advantage over
present-day `exact' non-LTE computations: all species responsible for metal
line blocking and blanketing can be considered, though approximately. The
`exact' non-LTE methods, on the other hand, are constrained to several abundant 
light elements and typically iron (and nickel), i.e. only the major 
line opacity sources are covered.
Note that these ODFs were calculated for solar abundances according to
Anders \& Grevesse~(\cite{AnGr89}). The latter have been revised in more recent
work such as Grevesse \& Sauval~(\cite{gre_sauv98}). In particular the
reduction of the abundance of iron (the most important line opacity source)
by $\sim$0.2\,dex should be considered. This is done in our work by
adopting the Kurucz~(\cite{kurucz93a}) ODFs for appropriately reduced
metallicity; see Sect.~\ref{sed} for further discussion.

\setcounter{table}{1}
\begin{table*}
\caption[]{Parameters of the programme stars \\[-6mm] \label{parameters}}
\begin{tabular}{lr@{$\pm $}lr@{$\pm $}lr@{$\pm $}lr@{$\pm $}lr@{$\pm $}lr@{$\pm $}l}
\hline
\noalign{\smallskip}
                 & \multicolumn{2}{c}{$\tau$\,Sco}   &
\multicolumn{2}{c}{HR\,3055}   & \multicolumn{2}{c}{HR\,1861}   &
\multicolumn{2}{c}{HR\,2928}   & \multicolumn{2}{c}{HR\,3468} & \multicolumn{2}{c}{HR\,5285}\\
\hline
$T_\mathrm{eff}$~(K)      &32\,000 & 300&31\,200 & 300&27\,000 & 400&
26\,300 & 400& 22\,900 & 400&21\,500 & 400\\
$\log g$~(cgs)     &4.30 & 0.05&3.95 & 0.05&4.12 & 0.05& 4.15 & 0.05&3.60 & 0.05&4.20 & 0.05\\
$\xi$~(km~s$^{-1}$)   &5 & 1      &  8 & 1    & 3 & 1     & 3 & 1   & 5 & 1 & 4 & 1\\
$v\sin i$~(km~s$^{-1}$)& 4 & 2  &  29 & 4   & 12 & 1    & 14 &  1  & 11 & 3 &18 & 1\\
$\zeta$~(km~s$^{-1}$) & 4 & 2    &  37 & 8   & \multicolumn{2}{c}{\ldots}  &
20 &  2   & 20 & 2 & \multicolumn{2}{c}{\ldots}\\
$y$ (by number) &  0.089 & 0.01 & 0.080 & 0.01  &
0.089 & 0.01  &  0.089 & 0.01  & 0.089 & 0.01   & 0.089 & 0.01\\
\hline
\end{tabular}
\end{table*}

The resulting non-LTE populations are then used to compute realistic line
profiles with SURFACE. The same microturbulent velocity as in DETAIL 
and in the model structure computations with ATLAS9 is adopted.
In this step of the calculation detailed Stark-broadening
data are employed, as summarised in
Table~\ref{tablelines} (available in the online version).
All other important data relevant to line formation are also given there:
wavelengths, lower and upper levels involved in the transition, oscillator
strengths $\log gf$, their accuracies, and sources.

Note that, in typical non-LTE computations for OB stars,
microturbulence is only included in the final profile calculation. Our choice
is based on test calculations that indicate line-profile fits are improved
if microturbulence is also included in computing the level
populations. The net effect is a slight strengthening of the lines
(Fig.~\ref{det_mic}). However, the effect is far less pronounced than described by
McErlean et al.~(\cite{McErleanetal98}), who investigated unblanketed
non-LTE models for B-type supergiants at slightly higher microturbulence;
cf. their Figs.~3~\&~4.

Our hybrid non-LTE approach involving the codes AT\-LAS9, DETAIL, and SURFACE
(henceforth abbreviated ADS) is tested here for early B-type dwarfs and
giants, supplemented by LTE computations with ATLAS9 and SURFACE (AS).
Our methodology may be applied to a wider range of stellar parameters
(i.e. effective temperature $T_\mathrm{eff}$ and surface gravity $\log g$).
Line-blanketed, static, and plane-parallel LTE models provide an
even more realistic description of stellar atmospheres at
lower temperatures and higher gravities (excited \ion{He}{ii} states should be
ignored at lower $T_\mathrm{eff}$ in order to avoid numerical inconsistencies).
Slightly higher temperatures (late O-types) and lower surface gravities
(less-luminous supergiants) may also be covered, until the hybrid non-LTE approach
meets its limitations when non-LTE effects on the atmospheric structure
and/or hydrodynamic mass-outflow may no longer be neglected.

\section{Observational data}

We test our analysis technique on six bright Galactic objects in the entire optical range
and also for near-IR lines when available. The programme stars
sample the parameter space in effective temperature and surface
gravity covered by typical applications.

High-S/N Echelle spectra of $\tau$\,Sco
 (\object{HR\,6165}), \object{HR\,3055}, \object{HR\,1861}, \object{HR\,2928}, \object{HR\,3468}, and
 \object{HR\,5285} were obtained by M.~Altmann
using FEROS (Fiberfed Extended Range Optical Spectrograph,
Kaufer et al.~\cite{Kauferetal99}) on the ESO 2.2m telescope in La Silla.
The data reduction was performed within the FEROS context
in the ESO MIDAS package, using optimum extraction.
The spectra were normalised by fitting a spline function to
carefully selected continuum points. This suffices to retain the line profiles
of the Balmer lines in these early-type stars. Finally, the spectra were
brought to the wavelength rest frame by cross-correlation with an
appropriate synthetic spectrum. Of the entire wavelength range covered by FEROS,
only the part between $\sim$3800 and 8000\,{\AA} meets our quality criteria
for further analysis. The spectra are compromised by the lower sensitivity of the instrument
at shorter wavelengths, and the reduced stellar fluxes in the far red.
FEROS provides a resolving power
$R$\,$\simeq$\,$\lambda/\Delta\lambda$\,$\approx$\,48\,000, with 2.2 pixels per
$\Delta\lambda$ resolution element. An S/N of up to $\sim$800 is achieved in
$B$. With respect to resolution and signal-to-noise ratio, the spectra available to us are 
of much higher quality than in typical studies of OB stars, basically
excluding observational uncertainties from the error budget.

\begin{figure}[h!]
\centering
\includegraphics[width=0.7\linewidth]{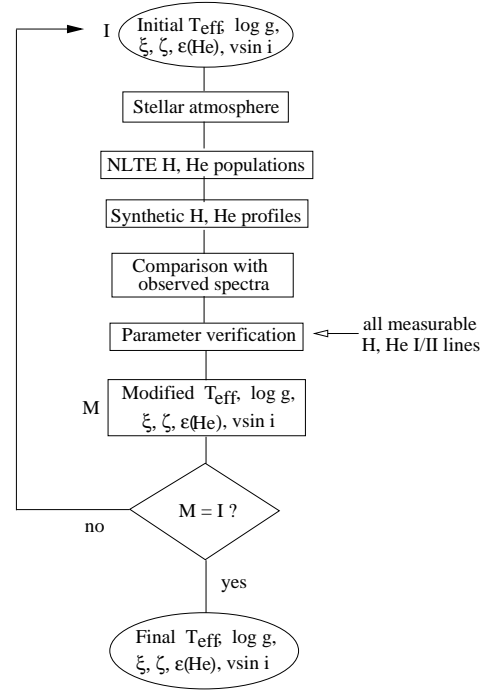}
\caption{Scheme of the iterative procedure for obtaining a simultaneous fit
to the hydrogen and helium lines.}
\label{diagram}
\end{figure}

A supplementary high-S/N spectrum of HR\,1861, also covering the higher Paschen
series, was obtained using FOCES (Fibre Optics Cassegrain Echelle
Spectrograph, Pfeiffer et al.~\cite{Pfeifferetal98}) on the
Calar Alto 2.2m telescope. The data were processed in a standard way,
using the data reduction routines described by Pfeiffer et al.~(\cite{Pfeifferetal98}).
An $R$\,$\simeq$\,40\,000 (2 pixels per
$\Delta\lambda$ resolution element) was achieved. In addition, a high-S/N  spectrum in the $K$-band  
of $\tau$\,Sco taken with Subaru/IRCS ($R$\,$\simeq$\,12\,000) is available
for analysis; see Hanson et al.~(\cite{Hansonetal05}) for details on the
observations and data reduction. Finally, a high-S/N spectrum in the 
$\lambda$ 2.058\,$\mu$m region of $\tau$\,Sco taken with UKIRT/CGS4 
($R$\,$\simeq$\,19\,000) is also available for analysis (Zaal et al.~\cite{zaal99}).

\section{Applications to observations}\label{obs}

\begin{figure}[h!]
\centering
\includegraphics[width=0.99\linewidth]{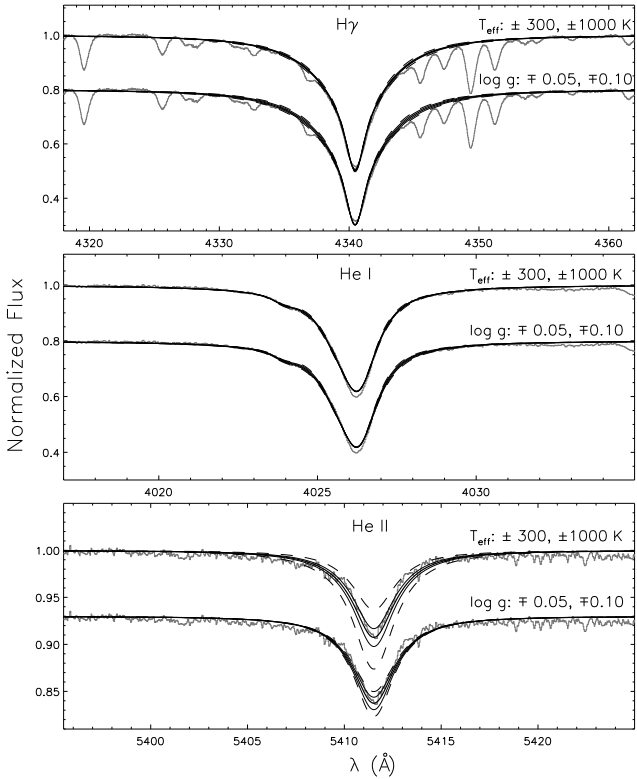}
\caption{ Impact of stellar parameter variations on non-LTE profile fits,
exemplarily for H$\gamma$, \ion{He}{i}\,$\lambda$4026\,{\AA}, and
\ion{He}{ii}\,$\lambda$5411\,{\AA} in HR\,3055 (B0\,III).
Synthetic spectra for our final parameters (see Table~\ref{parameters},
thick line) and varied parameters (thin
lines for our uncertainty estimates and dashed lines for values typically found in the literature) 
are compared to observation. See the text for a discussion. }
\label{uncertainties}
\end{figure}

Theoretical profiles were fitted to observations in an iterative
procedure summarised in Fig.~\ref{diagram}. The final atmospheric parameters
$T_\mathrm{eff}$ and $\log g$,
projected rotational velocities $v \sin i$, and micro- and
macroturbulent velocities ($\xi$, $\zeta$) coincide with those derived in Nieva \&
Przy\-billa~(\cite{np06a}), which are further refined by use of the
\ion{C}{ii/iii} and \ion{C}{ii/iii/iv} ionization equilibria (Nieva \&
Przy\-billa~\cite{np06b},c), see Table~\ref{parameters}.
The impact of stellar parameter variations on non-LTE profile fits to 
H$\gamma$, \ion{He}{i}\,$\lambda$4026\,{\AA}, and \ion{He}{ii}\,$\lambda$5411\,{\AA} 
in the hot giant HR\,3055 is visualised in Fig.~\ref{uncertainties}. Two
values for the parameter variations are adopted, according to our
uncertainties of 300\,K/0.05\,dex in $T_\mathrm{eff}$/$\log g$ and typical
values from the more recent literature (1000\,K/0.10\,dex). All
other hydrogen Balmer and helium lines react in a similar way. 
The sensitivity of the hydrogen and \ion{He}{i} lines to parameter variations
is low, such that the uncertainties cannot 
be reduced much below the typical values  even for high-S/N observations like ours. Only the \ion{He}{ii} lines are
highly sensitive to changes in $T_\mathrm{eff}$ and, to a lesser degree, in $\log g$. 
However, by taking metal ionization equilibria into consideration (e.g.
\ion{C}{ii/iii} or \ion{C}{ii/iii/iv}), which are even
more sensitive than \ion{He}{ii} lines, the parameters can be constrained more accurately. 

Projected rotational velocities, micro- and
macroturbulence values have also been verified by fitting the carbon lines.
Note that the comparatively high macroturbulence in HR\,3055 amounts to less
than twice the sound speed in the atmospheric plasma. The macroturbulent
broadening may thus be explained by a pattern of ascending and descending 
surface elements caused by (higher-order) nonradial
oscillations (see e.g. Lucy~\cite{Lucy76}).  
The microturbulence values are typically lower than found in previous work (e.g.
Kilian~\cite{Kilian92}). Differences in $T_\mathrm{eff}$ and $\log g$ are
also found.
Solar helium abundances $y$ (by number) are found in all cases.

\subsection{Visual}
\begin{figure*}[th!]
\centering
\includegraphics[width=0.995\linewidth]{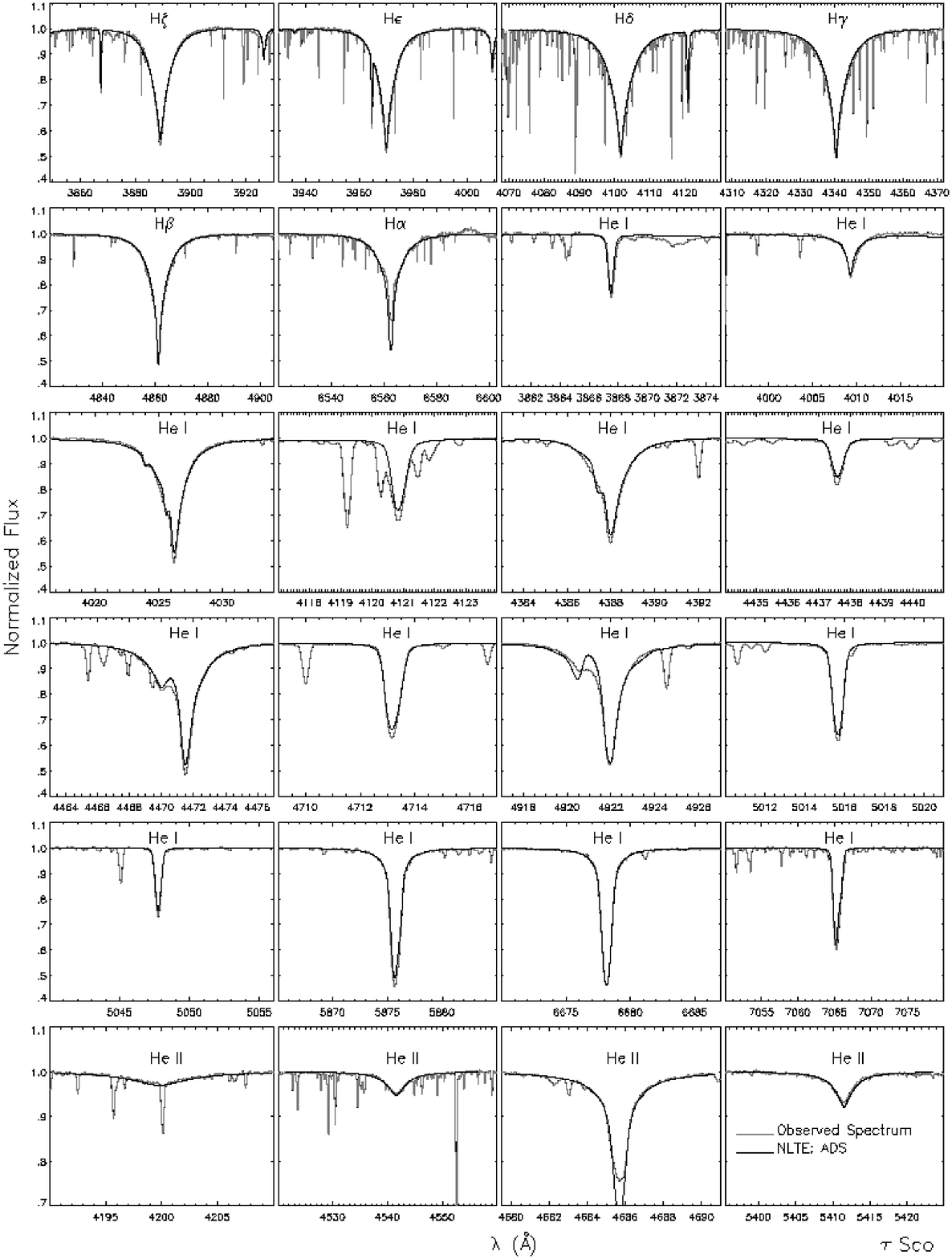}\\[-2mm]
\caption{Non-LTE line fits to observed hydrogen and helium features in
$\tau$ Sco (B0.2\,V), based on the atmospheric parameters
summarised in Table~\ref{parameters}.
}
\label{visualTsco}
\end{figure*}

\begin{figure*}[th!]
\centering
\includegraphics[width=0.995\linewidth]{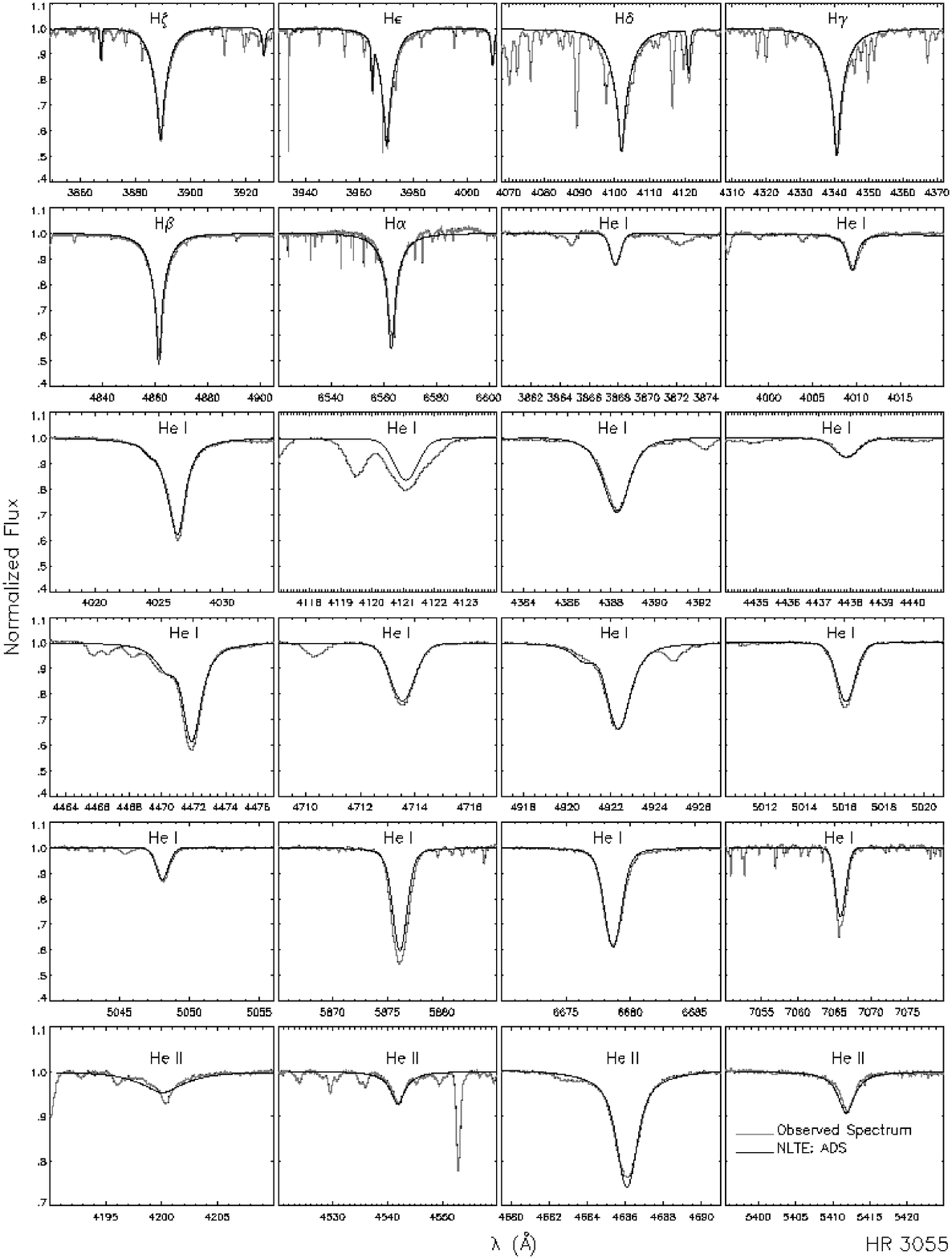}\\[-2mm]
\caption{Non-LTE line fits to observed hydrogen and helium features
in HR\,3055 (B0\,III). Note that the quality of the line fits for H$\alpha$ and
\ion{He}{ii}\,$\lambda$\,4686\,{\AA} in particular is better than for $\tau$ Sco
apparently because of a weaker stellar wind, cf. Fig.~\ref{visualTsco}.}
\label{visualHR3055}
\end{figure*}

Synthetic profiles for a selection of 6 hydrogen Balmer and 18
\ion{He}{i/ii} lines in the visual are compared with observation for the
sample stars in Figs.~\ref{visualTsco} and~\ref{visualHR3055} and in
Figs.~\ref{visualHR1861}--\ref{visualHR5285} (available in the online edition).
 They constitute our best simultaneous fits 
to the measurable H and He lines in the available spectra.
Preference for our selection has been given to (mostly) unblended features with 
good broadening data.
A summary of all available lines is given in Table~\ref{tablelines}, where
blending species are also identified. Our hybrid non-LTE approach (ADS) allows
us to reproduce the hydrogen Balmer and \ion{He}{i/ii} lines in the visual
more precisely, with few (well-understood) exceptions. The ionization equilibrium of
\ion{He}{i/ii} puts tight constraints on $T_\mathrm{eff}$ in the two hotter stars.

For the hottest dwarf of our sample, $\tau$\,Sco, a very good match between
model and observation is achieved (Fig.~\ref{visualTsco}),
except for the cores of H$\alpha$ and \ion{He}{ii}
$\lambda$4686\,{\AA}. This is because of the neglect of the stellar
wind; see Przybilla \& Butler~(\cite{przyb-but04}) and Mokiem et
al.~(\cite{Mokiemetal05}) for results of hydrodynamic computations.
The discrepancies in \ion{He}{i} $\lambda$4121\,{\AA} occur because of
blends with metal lines (\ion{O}{ii}, \ion{C}{iii}, and \ion{Fe}{iii}; unaccounted for in our computations), which can
be nicely resolved at this low $v \sin i$. An improved fit to \ion{He}{i}
$\lambda$4921\,{\AA} may be obtained with better broadening data for the
forbidden component.The spectral region around H$\alpha$ suffers from artifacts introduced by CCD
defects that can only partially be compensated for in the data reduction process.

Line fits to the hot giant HR\,3055 are displayed in Fig.~\ref{visualHR3055}.
Excellent agreement between theory and observation is also found in this
case, with a significantly improved fit quality of H$\alpha$ and \ion{He}{ii}
$\lambda$4686\,{\AA}, because of an apparently weaker wind.
This star shows a higher $v \sin i$ and $\zeta$
than $\tau$\,Sco. Therefore the \ion{He}{i} $\lambda$4121\,{\AA} blend
is no longer resolved, leading to an apparently worse fit.

\ion{He}{ii} $\lambda$4686\,{\AA} is the only visible (weak) feature of
\ion{He}{ii} in the intermediate temperature stars HR\,1861 and HR\,2928;
see Figs.~\ref{visualHR1861} and~\ref{visualHR2928}.
Good fits are obtained for this line and the features of
\ion{He}{i} (i.e. establishing the ionization equilibrium) and hydrogen.
Lines of \ion{He}{ii} are absent at even lower temperatures, HR\,3468 and
HR\,5285 in our sample, see Figs.~\ref{visualHR3468} and~\ref{visualHR5285}.

\subsection{Near-IR}
\begin{figure}
\centering
\includegraphics[width=\linewidth]{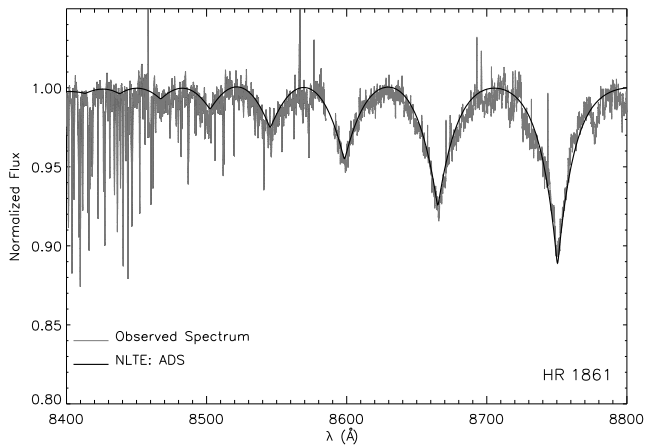}
\caption{Modelling of the Paschen series of HR\,1861 with our non-LTE
(ADS) approach. Note the presence of numerous sharp telluric
H$_2$O lines.  All theoretical spectra in the near-IR have been
computed with the same atmospheric parameters than the models in the visual.
}
\label{paschen}
\end{figure}

\begin{figure}
\centering
\includegraphics[width=\linewidth]{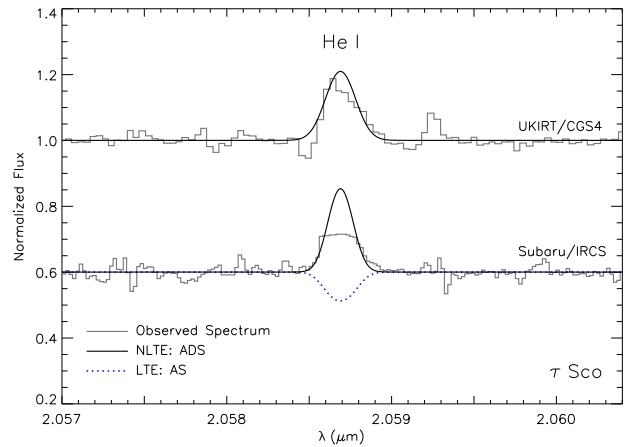}
\caption{ Modelling of the \ion{He}{i} $\lambda$2.058\,$\mu$m singlet line in $\tau$\,Sco. 
}
\label{he1k1}
\end{figure}

\begin{figure}
\centering
\includegraphics[width=\linewidth]{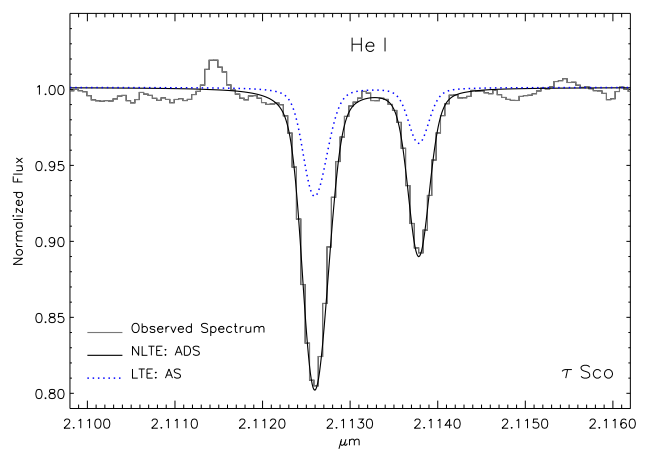}
\caption{Modelling of the \ion{He}{i} $\lambda$2.11\,$\mu$m singlet and
triplet feature in $\tau$\,Sco. As also shown in Fig.~\ref{he1k1}, these near-IR 
transitions experience stronger non-LTE effects than the spectral lines in the visual.}
\label{he1k}
\end{figure}

Additional spectra are available in the near-IR for two stars. An excellent fit
to the higher Paschen series is obtained for HR\,1861 in non-LTE, despite the relatively low
S/N and the presence of telluric lines, see Fig.~\ref{paschen}.
 Good agreement between the non-LTE spectrum synthesis
and observation can also be obtained for the \ion{He}{i}
$\lambda$ 2.058\,$\mu$m feature and 
practically perfect agreement for the $\lambda$2.11\,$\mu$m
 lines in $\tau$\,Sco, see Figs.~\ref{he1k1} and~\ref{he1k}.
 Note that the $\lambda$ 2.058\,$\mu$m transition is situated in an
atmospheric window with a series of strong telluric lines, therefore its shape is highly
sensitive to the detailed approach in the data reduction process
(see e.g. Zaal et al.~\cite{zaal99}). 
The two different observations in Fig.~\ref{he1k1} may exemplify the difficulty of accurate telluric line
removal in the data reduction process (or alternatively an intrinsic time
variability of the feature). 
Good agreement with our model is obtained in the case of the higher-resolution
UKIRT/CGS4 spectrum. The LTE approach is not even able to reproduce the observation qualitatively.
The line fits in the near-IR are based on the same atmospheric parameters 
(Table~\ref{parameters}) as used for the modelling of the optical spectra.

Non-LTE effects can be amplified in the Rayleigh-Jeans part of the
spectral energy distribution, as demonstrated in this case. See e.g.
Przybilla \& Butler~(\cite{przyb-but04}) for a discussion and for line
fits to Br$\gamma$ in the $K$-band and to additional Brackett and Pfund lines
in this star. The case of \ion{He}{i} $\lambda$10\,830\,{\AA} has been
discussed by Przybilla~(\cite{przyb05}).

\section{Comparison with other model predictions}
In this section we compare our hybrid non-LTE computations (ADS) with four other approaches. 
Two of them are available grids from the literature, and to understand their 
discrepancies to our ADS computations we calculate additional non-LTE and LTE models.

{\sc i}) We replace the ATLAS9 models by line-blanketed, plane-parallel, and hydrostatic 
non-LTE model atmospheres taken from the publically available OSTAR2002 grid (Lanz \&
Hubeny~\cite{la-hu03}, LH03) and carry out the non-LTE line formation calculation with 
DETAIL and SURFACE as described above.

{\sc ii}) We use the non-LTE calculations from the OSTAR2002 grid (TLUSTY and non-LTE line
 formation with SYNSPEC), published by LH03.

{\sc iii}) We calculate LTE spectra based on ATLAS9 atmospheres and
 subsequent LTE spectrum synthesis with SURFACE (AS).

{\sc iv}) We use the Padova grid (Munari et al.~\cite{mu03}), based on ATLAS9
atmospheres and LTE spectrum synthesis performed with the
SYNTHE code (Kurucz \& Avrett~\cite{kav81}; Kurucz~\cite{kurucz93c}).

\subsection{Atmospheric structures, SEDs: LTE vs. non-LTE}\label{sed}
\begin{figure*}
 \centering
  \includegraphics[width=.85\linewidth]{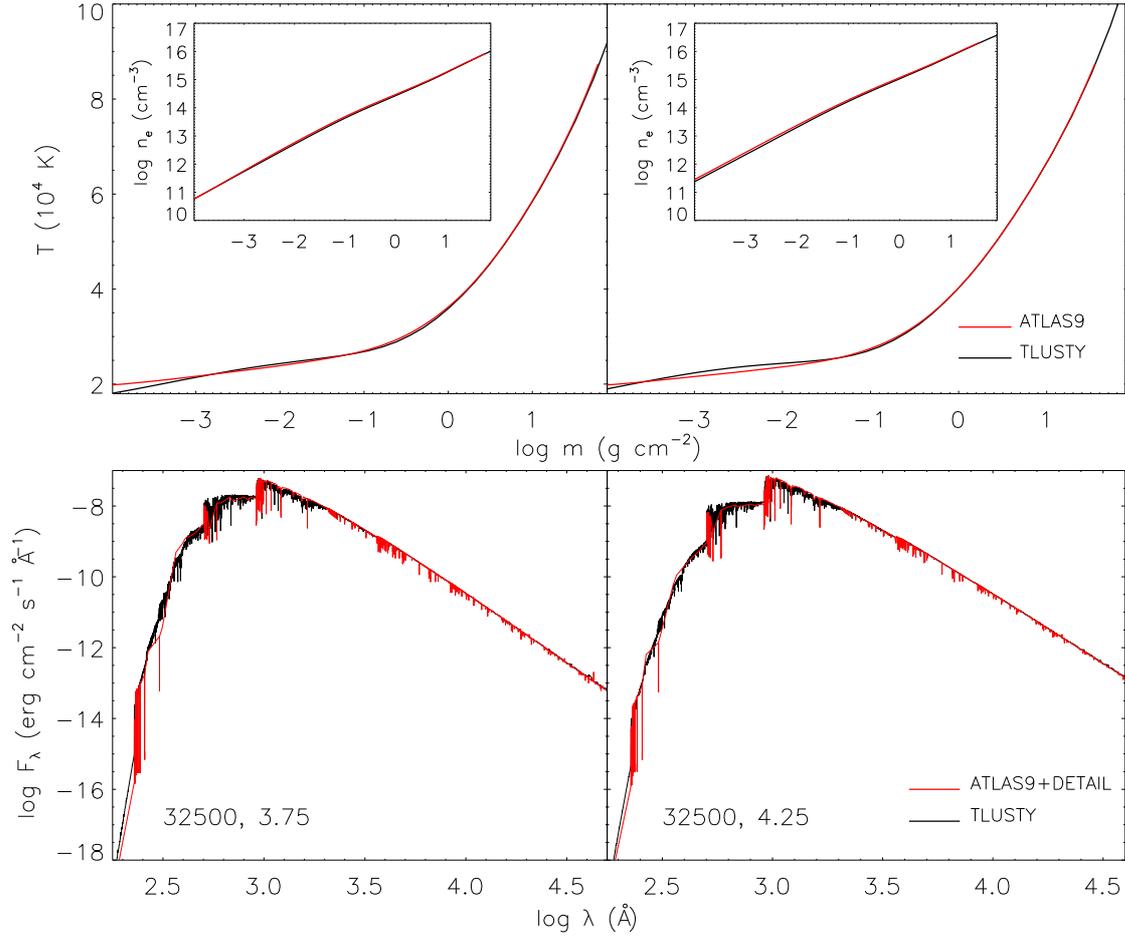}\\[-4mm]
\caption{Upper panel: Comparison of AT\-LAS9 and TLUSTY temperature structures and electron
densities (insets) as func\-tion of column mass. The computations have
been performed for giant and dwarf models.
Lower panel: Comparison of spectral energy distributions, the
radiation field computed by DETAIL on the basis of the ATLAS9 atmospheric
structure vs. TLUSTY. 
}
\label{str_flux}
\end{figure*}

A comparison of LTE (ATLAS9) and non-LTE (TLUSTY) atmospheric structures and spectral energy
distributions (SEDs) computed with ATLAS9$+$DETAIL and TLUSTY
is made in Fig.~\ref{str_flux}. Models for a hot giant and a dwarf
($T_\mathrm{eff}$\,$=$\,32\,500\,K, $\log g$\,$=$\,3.75 and 4.25, respectively) are considered,
approximately delineating the upper temperature boundary of our observations
($\tau$\,Sco). Reduced non-LTE effects can be expected for cooler models.

\begin{figure}
\centering
\includegraphics[width=.98\linewidth]{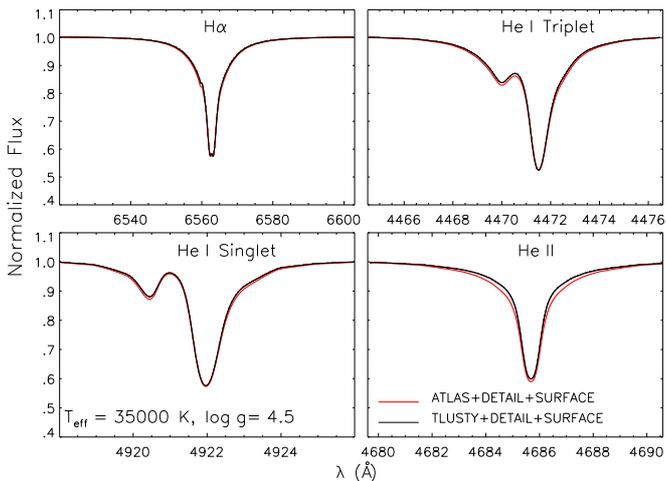}
\caption{Comparison of the most discrepant hydrogen and \ion{He}{i/ii} line profiles from our hybrid
non-LTE approach (ADS) and a TLUSTY-DS calculation for a hot 
main-sequence model. Practically perfect agreement is obtained, with small
discrepancies occurring only in the wings of \ion{He}{ii} $\lambda$4686\,{\AA}.}
\label{ADS_TDS}
\end{figure}

Excellent agreement is found for the temperature and density structures.
This is a basic requirement for successful application of the hybrid non-LTE approach
for spectrum synthesis.
The temperature structures deviate by less than 1\% in the inner
atmosphere, including the regions where the weaker lines and the wings
of the stronger features are formed ($\log m$\,$\gtrsim$\,$-$1; see Sect.~\ref{linfor}).
At the formation depths of the cores of the stronger
H and He lines ($-$3\,$\lesssim$\,$\log m$\,$\lesssim$\,$-$1.5) the
differences may increase to $\lesssim$\,2--3\%. Stronger
deviations may occur only in the outermost
parts of the atmosphere, outside the line-formation depths.
Note that this good a match is obtained only if the effects of
metal line-blanketing are correctly accounted for. In particular, the
computations should be made for identical metal abundances. This is
complicated by the fact that the ODFs of Kurucz~(\cite{kurucz93a}) were
computed assuming scaled solar abundances from Anders \&
Grevesse~(\cite{AnGr89}), while the TLUSTY computations assume 
abundances from Grevesse \& Sauval~(\cite{gre_sauv98}). The most important
difference is a downward revision of the iron abundance by $\sim$0.2\,dex in the
later work. Consequently, we use ODFs with correspondingly `sub-solar'
metallicity in order to correct for the discrepancies in the line opacities,
with $[$Fe/H$]$ as a metallicity substitute.
See also Przybilla et al.~(\cite{Przybillaetal06b}) for a discussion of such
`empirical' corrections.
We should note that while the differences are small at (near-)solar
abundances the non-LTE effects on the atmospheric structure will increase
with decreasing metallicity. Nevertheless, our hybrid non-LTE methodology for OB star
 analyses should be applicable down to SMC metallicity, as
indicated by an analogous comparison for models at 1/5\,$\times$\,solar abundances.

\begin{figure*}
\centering
\includegraphics[width=\linewidth,height=13.8cm]{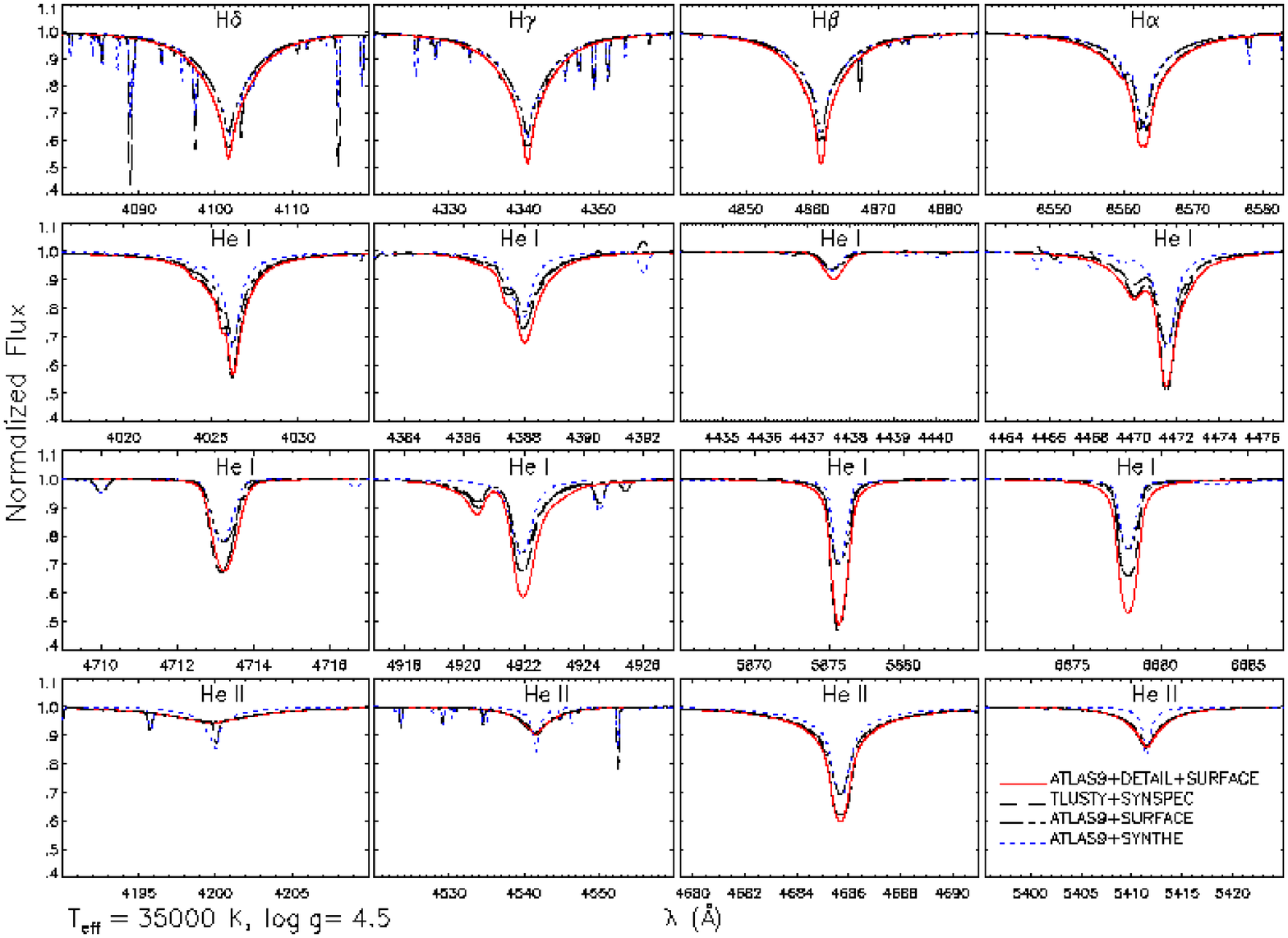}
\caption{Comparison of selected H and \ion{He}{i/ii} line profiles from our hybrid
non-LTE approach (ADS), non-LTE computations from TLUSTY$+$SYNSPEC, and two LTE calculations
(ATLAS9$+$SURFACE and ATLAS9$+$SYNTHE) for a hot 
main-sequence model. 
}
\label{ADSvsall}
\end{figure*}

The SEDs computed with ATLAS9+DETAIL and TLUSTY show excellent agreement over almost
the entire wavelength range. Small differences occur in the EUV, most
notably in the \ion{He}{ii} continuum. This is a significant improvement over
the comparison of TLUSTY with ATLAS9 model fluxes (not shown here), which
predict much lower fluxes in the Lyman and helium continua. The LTE computations
neglect the non-LTE overionization of the
hydrogen and \ion{He}{i} ground states. This overionization reduces the bound-free
continuum opacity; see Sect.~\ref{linfor} for a more comprehensive discussion.

\subsection{Spectra: hybrid non-LTE vs. full non-LTE and
LTE}\label{comparison}

\begin{figure*}
\centering
\includegraphics[width=\linewidth,height=14.3cm]{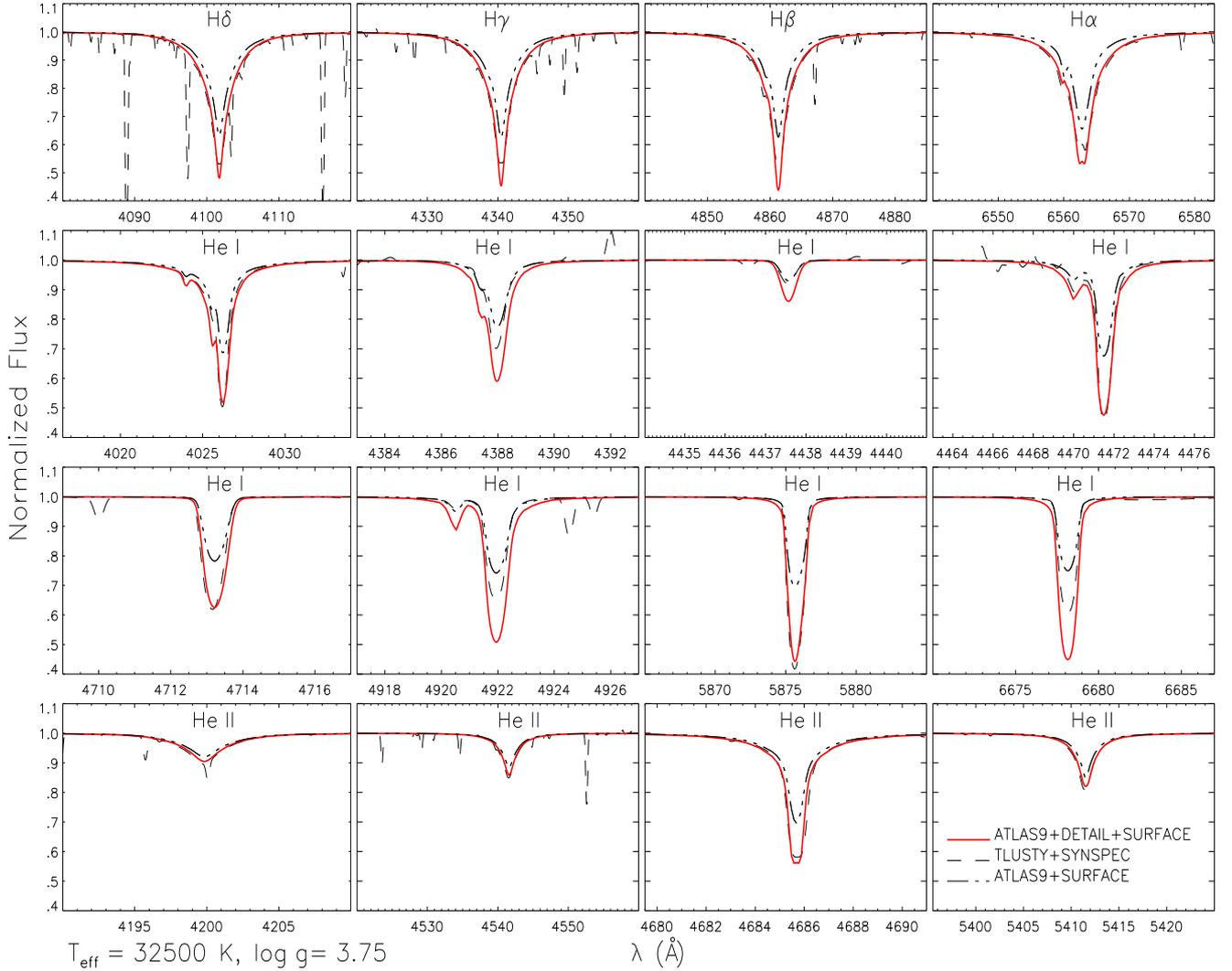}
\caption{Comparison of selected H and \ion{He}{i/ii} line profiles from our
hybrid non-LTE approach (ADS), the non-LTE computations with TLUSTY$+$SYNSPEC and our
LTE approach ATLAS9$+$SURFACE for a hot giant model. Here, HR\,3055 may act as observational discriminator, indicating our
results to be appropriate (see Fig.~\ref{visualHR3055}).
}
\label{ADSvsfNLTE}
\end{figure*}

\begin{figure*}
\centering
\includegraphics[width=\linewidth,height=12.7cm]{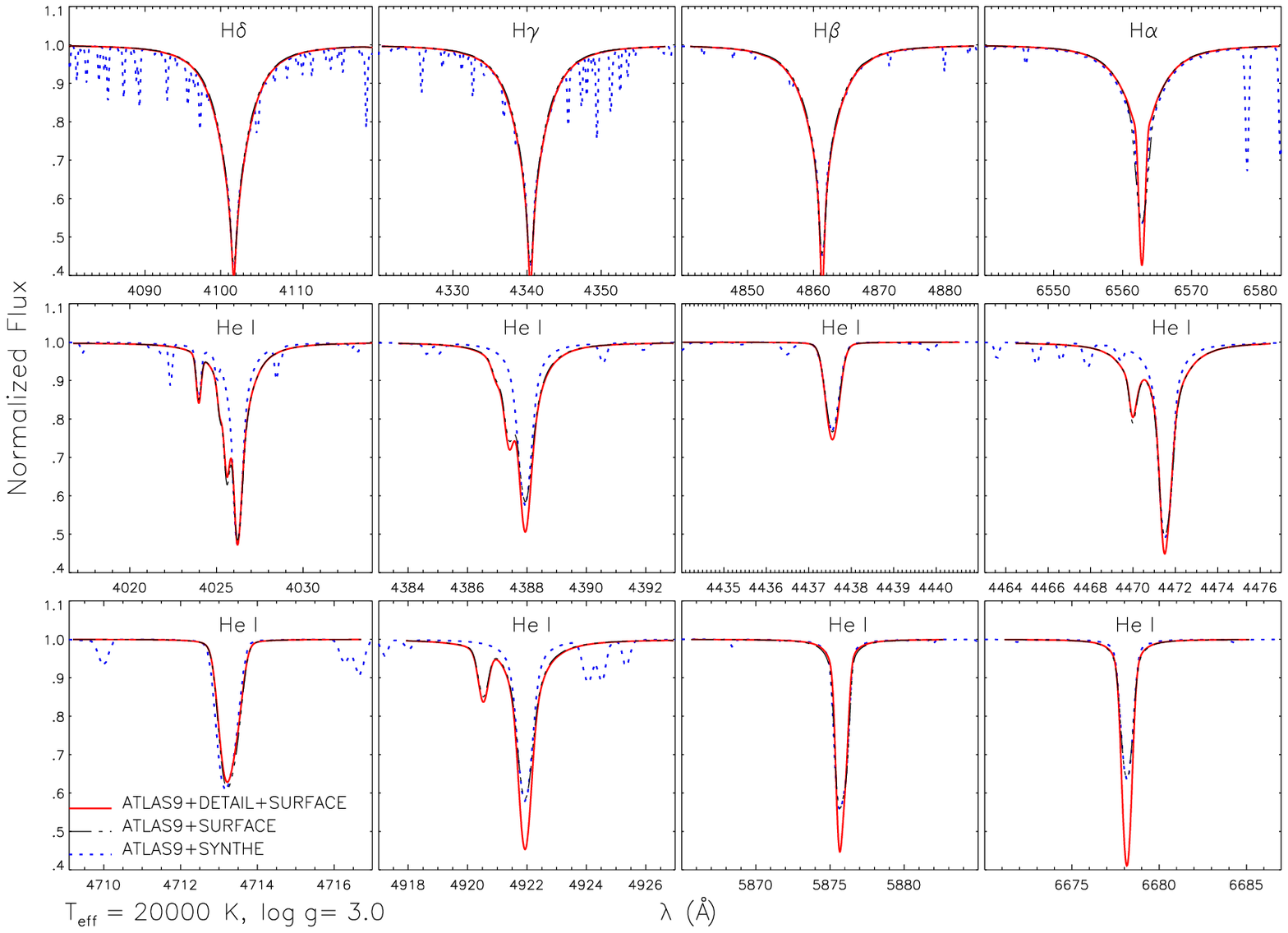}
\caption{H and \ion{He}{i} profiles for a cool giant model: our hybrid non-LTE
approach (ADS) vs. our LTE (ATLAS9$+$SURFACE) and the corresponding Padova model (ATLAS9$+$SYNTHE);
see Fig.~\ref{visualHR3468} for the closest observational analogue. 
}
\label{ADSvsfLTE}
\end{figure*}

Comparisons of synthetic profiles of several strategic lines of hydrogen and
\ion{He}{i/ii} are made for three test cases, where models are
available from the published grids. These frame the parameter space
studied in the present paper. The test cases comprise a hot dwarf model
($T_\mathrm{eff}$\,$=$\,35\,000\,K, $\log g$\,$=$\,4.5), at slightly higher
$T_\mathrm{eff}$ than covered by our observations\footnote{We expect our
approach to be valid at slightly hotter temperatures, at least in main
sequence stars where the stellar wind does not influence the photospheric
layers strongly.}, shown in Fig.~\ref{ADSvsall};
 a hot giant model (32\,500, 3.75), with similar temperature to
$\tau$\,Sco, see Fig.~\ref{ADSvsfNLTE}; a cool giant model
(20\,000, 3.00), with both $T_\mathrm{eff}$ and $\log g$ lower than covered
by our observations, Fig.~\ref{ADSvsfLTE}. In order to be consistent with the published
grids, our computations consider solar metal abundances (Grevesse \&
Sauval~\cite{gre_sauv98}) and solar helium abundance. Our synthetic
spectra and those of the OSTAR2002 grid are degraded to the highest resolution
($R$\,$=$\,20\,000) available from the Padova grid. Note that metal lines are neglected when 
we calculate the emergent spectrum, and they are considered indirectly via line blanketing/blocking effects.

{\sc i}) ADS vs. TLUSTY-DS. This comparison allows effects caused
by differences in the model atmosphere structures to be disentangled.
A practically perfect match for the (35\,000,4.5) model (see Fig.~\ref{ADS_TDS}) is obtained in the
ADS and TLUSTY-DS computations, which share the same model atom. This indicates
good agreement of the LTE and non-LTE atmospheric structures at even slightly higher
temperatures than discussed in Fig.~\ref{str_flux}. We refrain
from presenting further comparisons of these two approaches, as the discrepancies are
even smaller at lower temperatures.

{\sc ii}) ADS vs. TLUSTY$+$SYNSPEC (LH03). These results are obtained using two independent methods
(model atmospheres, model atoms, numerical solution). Nonetheless, good
agreement is found on the whole for the (35\,000,4.5) and (32\,500,3.75) models
from an inspection of Figs.~\ref{ADSvsall} and~\ref{ADSvsfNLTE}, respectively.
Notable differences between ADS and TLUSTY$+$SYNSPEC
occur in the line cores of the Balmer lines (the latter filled in by
emission) and in the \ion{He}{i} singlet lines, which are systematically
weaker in the case of TLUSTY$+$SYNSPEC, in contrast to observation (see Figs.~\ref{visualTsco} and \ref{visualHR3055}). 
The \ion{He}{i} triplet lines derived from both approaches agree well.
Small discrepancies occur in the line wings of the \ion{He}{i} lines because
of different broadening data. A good match is also obtained for the
\ion{He}{ii} lines, with small differences arising in \ion{He}{ii}
$\lambda$4686\,{\AA}.

From the comparison of the TLUSTY$+$SYNSPEC and TLUSTY-DS results, which match our ADS, 
we can conclude that the aforementioned discrepancies arise because of subtle differences in
solving the statistical equilibrium and radiative transfer problem. While
our approach uses line opacities averaged over the ODF wavelength bins, the
TLUSTY$+$SYNSPEC computations employ a more sophisticated opacity
sampling technique. This however introduces a strong dependency on the model
assumptions for \ion{Fe}{iv}\footnote{In particular 
on the oscillator strengths of the \ion{Fe}{iv} transitions involved. A 
reduction of the $gf$-values may alleviate the discrepancy between the
\ion{He}{i} singlet and triplet line strengths, seen, for example, in the
OSTAR2002 grid.}
(a highly complex ion), which has lines overlapping with an \ion{He}{i} resonance
transition 
(Najarro et al.~\cite{Najarroetal06})\footnote {As a further
test we have calculated synthetic spectra for the (35\,000,4.5) and
(32\,500,3.75) models on the 
basis of unified, line-blanketed non-LTE model atmospheres (FASTWIND, Puls et
al.~\cite{Pulsetal05}). Excellent agreement with our ADS results is found for 
both the \ion{He}{i} singlet and triplet lines. 
Note that FASTWIND also uses an approximate
treatment of line blocking.}. 
The same model atmosphere (TLUSTY) is
used and both model atoms should be sufficiently robust for modelling the lines in the
visual; see Przybilla \& Butler~(\cite{przyb-but04}) and Przybilla~(\cite{przyb05})
for a discussion of this.

{\sc iii}) ADS vs. ATLAS9$+$SURFACE.
LTE computations with AS produce narrower Balmer lines for the (35\,000,4.5)
and (32\,500,3.75) models (the
differences reducing progressively from H$\alpha$ to the higher series
members), which leads to overestimated surface gravities in that case. At
the same time, {\em all} \ion{He}{i} lines are too shallow in LTE, the trend
increasing from blue to red and showing larger discrepancies at lower gravity. On the other hand, rather
good agreement is found for the \ion{He}{ii} lines, the LTE predictions being slightly
weaker than ADS for the hot giant.
In Fig.~\ref{ADSvsfLTE} a comparison of our hybrid non-LTE with our pure LTE prediction
is made for a (20\,000,3.0) model with
a temperature slightly below than the lower limit of our programme stars, and at significantly reduced surface gravity.
Here, the wings of the Balmer lines show much
better agreement than at higher temperatures (cf. Fig.~\ref{ADSvsall}), as well as
 the \ion{He}{i} $\lambda\lambda$ 4437 and 4713\,{\AA} lines. 
The
line cores are also discrepant, increasingly so from H$\delta$ to H$\alpha$. Many of the
\ion{He}{i} lines experience significant non-LTE strengthening, in
particular those in the red. The line broadening data is the same in ADS and AS,
so the \ion{He}{i} wings are very similar. The forbidden components are also accounted
for in both approaches.

{\sc iv}) ADS vs. ATLAS9$+$SYNTHE (Munari et al.~\cite{mu03}). The differences
of these approaches were quantified for the (35\,000,4.5) model, when possible.
The Balmer lines from the Padova model present similar characteristics as the LTE AS approach
(in Fig.~\ref{ADSvsall} they coincide), resulting in lower equivalent widths by up to $\sim$30\% 
relative to ADS.
When using the H$\gamma$ wings as a surface-gravity indicator, this translates to a systematic 
error in $\log g$ by $\sim$0.2\,dex, implying even larger errors for fits to the H$\beta$ and
H$\alpha$ wings.
The \ion{He}{i} lines are generally too weak, by up to a factor of more than~2 in equivalent width, and the \ion{He}{ii} lines too narrow.
For the most part, these discrepancies stem from the neglect of non-LTE
effects on the line-formation process, as the differences in the atmospheric
structures are practically insignificant. 

\begin{figure*}
\includegraphics[width=.90\linewidth]{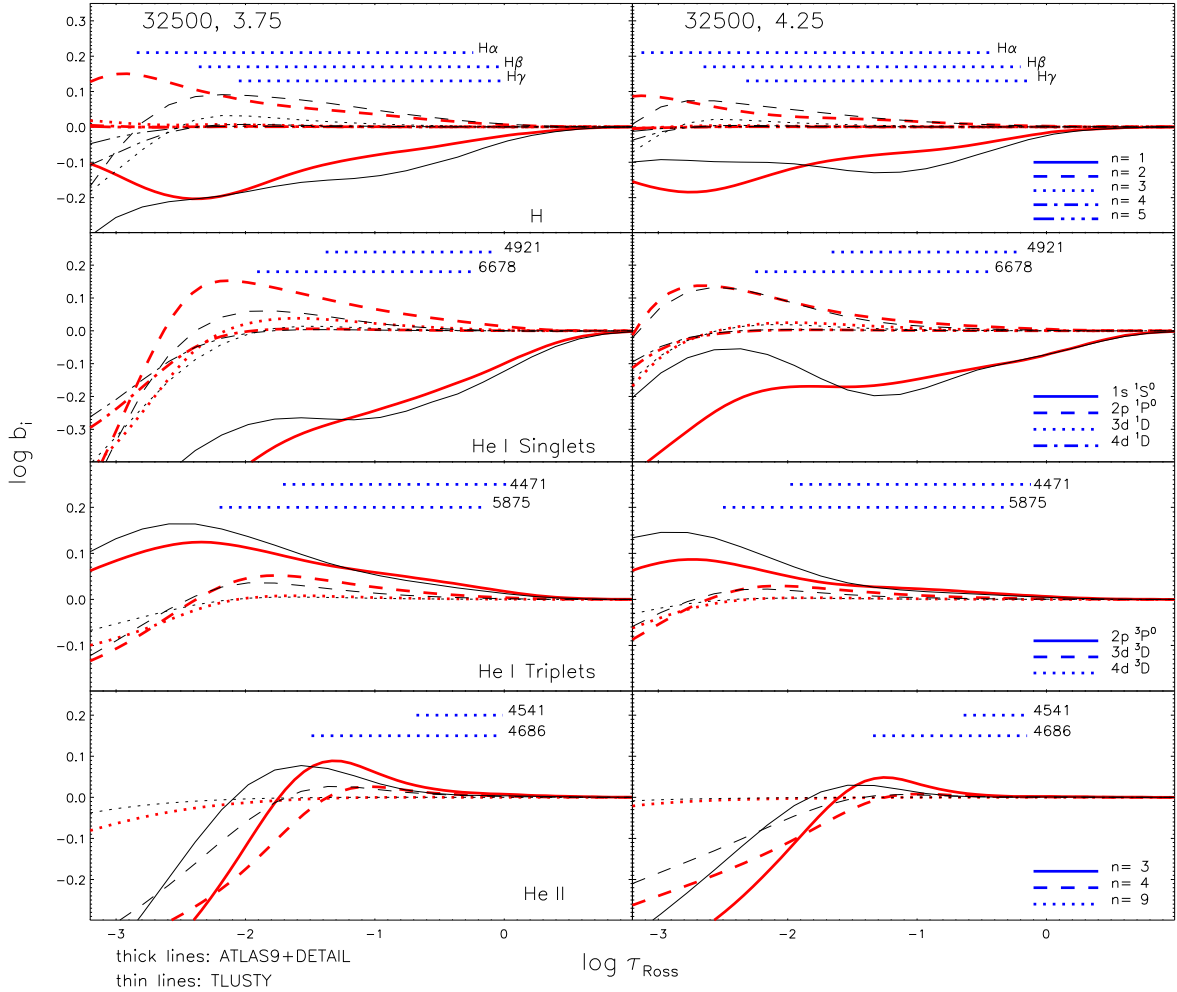}
\caption{Departure coefficients $b_i$ of some strategic hydrogen and helium
levels as a function
of Rosseland optical depth $\tau_\mathrm{Ross}$. The comparison is made
for the giant (left) and dwarf atmospheric models (right column) already
discussed in Fig.~\ref{str_flux}, for our approach (ATLAS9$+$DETAIL, thick lines) and the
results of LH03 (TLUSTY, thin lines). Each level is coded by different line styles;
see the legend in the corresponding panels. Line-formation regions (from
core to wing) corresponding to our calculations are indicated. See the text
for further discussion.}
\label{bi}
\end{figure*}

Another limiting factor of the
ATLAS9$+$SYNTHE computations is the use of insufficient Stark broadening
data (Voigt profile with constant Stark damping parameter). 
Our AS and ADS approaches improve 
on this, as realistic
broadening data is used
(see Table~\ref{tablelines}). In the ATLAS9$+$SYNTHE approach, it will not be possible to
obtain reasonable agreement for the \ion{He}{i} and \ion{He}{ii} spectra at the same time.
For the (20\,000,3.0) model, the Padova profiles are more similar to the AS approach.
 However, the diffuse \ion{He}{i} lines still suffer from inappropriate
broadening data,
in particular the forbidden components are unaccounted for.
The \ion{He}{i} lines are affected by non-LTE strengthening, increasing to the red.
Only few \ion{He}{i} lines are quite similar in the three
approaches:
$\lambda\lambda$4437 and 4713\,{\AA} match quite~well, as do 
$\lambda\lambda$3867, 4121 (despite blends with metallic lines) and 5015 and 5047\,{\AA}, 
not displayed here.\\[-3mm]

We should mention that the published libraries of synthetic spectra were
 computed with different values of microturbulent velocity
(OSTAR2002: 10\,km\,s$^{-1}$; Padova: 2\,km\,s$^{-1}$). The ADS and TLUSTY-DS calculations
with $\xi$\,$=$\,10\,km\,s$^{-1}$ were adopted for the comparison in
Fig.~\ref{ADSvsall}. We made tests with a reduced $\xi$\,$=$\,2\,km\,s$^{-1}$,
resulting in only small changes in the \ion{He}{ii} profiles -- the most
sensitive to modifications of $\xi$. The differences between the Padova grid and
the other approaches are indeed due to the neglect of non-LTE effects and to
additionally insufficient broadening data and not because of discrepant
microturbulences. The ADS and AS computations in Figs.~\ref{ADSvsfNLTE} and~\ref{ADSvsfLTE}
were performed using the $\xi$ of the respective libraries.

\subsection{Line formation: hybrid non-LTE vs. full non-LTE}\label{linfor}

\begin{figure*}
\includegraphics[width=.90\linewidth]{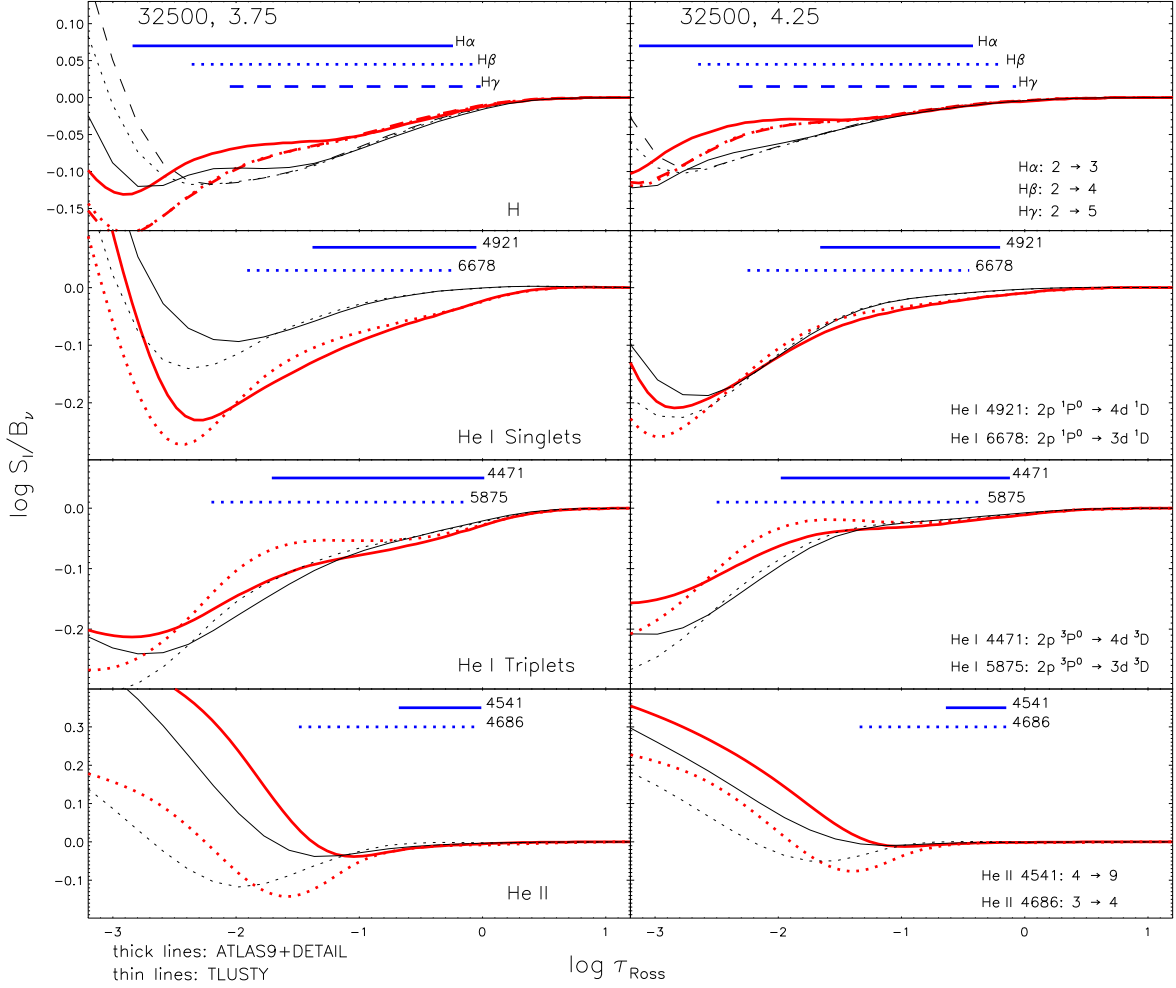}
\caption{Ratio of line source function $S_\mathrm{l}$ to Planck function $B_\nu$ at
line centre as a function of $\tau_\mathrm{Ross}$ for selected spectral lines
of hydrogen and helium.
The comparison is made in analogy to Fig.~\ref{bi}, for our approach
(ATLAS9$+$DETAIL, thick lines) and LH03 (TLUSTY, thin lines). The individual
spectral lines are encoded by the different line styles indicating the
line-formation depths (from our calculations). See the text
for further discussion.}
\label{sl}
\end{figure*}

We now try to identify the physical reasons for the differences in the non-LTE line 
profiles of hydrogen and helium in the
last comparisons by a closer study of the underlying
line-formation processes. For this we investigate non-LTE departure
coefficients and line source functions for three representative hydrogen and
six \ion{He}{i/ii} lines, as derived in our ATLAS9$+$DETAIL and the
TLUSTY computations. We choose the same models as discussed in
Fig.~\ref{str_flux}. For the (32\,500, 3.75) model, a direct comparison with the
resulting line profiles is facilitated by inspection of Fig.~\ref{ADSvsfNLTE}.

For the levels involved in the transitions of interest and the hydrogen and
helium ground
states, departure coefficients $b_i$ (referred to the ground
state of the next higher ion) are displayed in Fig.~\ref{bi}.
The non-LTE effects on the level occupations give rise to departures of the
line source function $S_\mathrm{l}$ from the Planck function $B_\nu$; see
Fig.~\ref{sl} for a comparison of $S_\mathrm{l}/B_\nu$ from the ATLAS9+DETAIL and
TLUSTY computations. We recall that
\begin{equation}
S_\mathrm{l} = \frac{2h\nu^3/c^2}{b_i/b_j \exp{(h \nu /kT)} - 1}~,
\end{equation}
with $h$ the Planck and $k$ the Boltzmann constant, $\nu$ the transition
frequency, and $T$ the local temperature.
For given $T_\mathrm{eff}$ the non-LTE effects are strengthened with
decreasing surface gravity, implying lower particle densities and thus
larger mean-free-paths between photon absorptions.

\paragraph{Hydrogen.} We study three hydrogen lines, H$\alpha$, H$\beta$
and H$\gamma$. Departure coefficients for levels $n$\,$>$\,5 behave similarly
to those for $n$\,$=$\,5, which already traces the behaviour of the continuum closely.
Consequently, the line source functions for the higher Balmer lines are similar
to that of H$\gamma$. The non-LTE depopulation of the H ground state (Fig.~\ref{bi})
 reduces the
Lyman continuum opacity, giving rise to higher EUV fluxes than in LTE.
The Lyman lines are expected to experience non-LTE weakening.
Note that the TLUSTY calculations indicate a slightly stronger non-LTE
depopulation of the ground state at continuum formation depths than in
our case. Good agreement of the departure coefficients for $n$\,$=$\,2 is
found, which is overpopulated at line formation depths. In combination with
the higher H states being close to LTE, this explains the non-LTE strengthening
of the Balmer lines. At the formation depths of the line cores, in
particular for H$\alpha$, the TLUSTY results show a less pronounced
overpopulation, eventually leading to an underpopulation of the $n$\,$=$\,2
state in the outer atmosphere. This explains the shallower lines from the
TLUSTY computations relative to ADS,
which is a consequence of the upturn of $S_\mathrm{l}/B_\nu$~(Fig.~\ref{sl}).
The effect decreases from H$\alpha$ to the higher Balmer lines, as the core
formation depths shift to deeper~atmospheric~layers.

\paragraph{\ion{He}{i} Singlets.} We investigate two representative features: $\lambda\lambda$4921 and 6678\,\AA. In general, the departure coefficients
for most of the levels show rather good agreement, in qualitative behaviour as well as
quantitatively. Notable differences in the ground-state overionization occur in
the outer atmosphere. More relevant is the behaviour of the
2p\,$^1$P$^{\rm o}$ level, the lower level of practically all \ion{He}{i}
singlet transitions in the visual. The non-LTE overpopulation in the TLUSTY
results is far less-pronounced than in our case (Fig.~\ref{bi}), in particular for the giant
model. This gives rise to discrepant line source functions (Fig.~\ref{sl}) and consequently
differing line profiles (Fig.~\ref{ADSvsfNLTE}) in the two
approaches, with the ADS computations correctly predicting the non-LTE
strengthening and thus reproducing observation (Sect.~\ref{obs}). The states
at higher excitation energies ($n$\,$\ge$\,4) are in LTE relative to
the \ion{He}{ii} ground state at line-formation depths.\\[-.6cm]

\paragraph{\ion{He}{i} Triplets.} We investigate two representative lines:
$\lambda\lambda$4471 and 5875\,\AA. The departure coefficients from the two
approaches differ only slightly at line-formation depths (Fig.~\ref{bi}).
As a consequence, the source functions (Fig.~\ref{sl}) are also similar, resulting
in negligible differences of the line profiles (Fig.~\ref{ADSvsfNLTE}).
Again, the states with $n$\,$\ge$\,4 are in detailed balance relative to the
\ion{He}{ii} ground state at depths relevant for the line formation.

\paragraph{\ion{He}{ii}.} Two features were studied: $\lambda\lambda$4541
and 4686\,\AA. The higher \ion{He}{ii} levels are close to LTE with
the continuum state at line-formation depths. Only the $n$\,$=$\,3 level shows a
relevant overpopulation, resulting in a non-LTE strengthening of the
$\lambda$4686\,{\AA} line. The differences in the ADS and TLUSTY departure
coefficients give slightly shallower profiles for this line in the
OSTAR2002 model (Fig.~\ref{ADSvsfNLTE}). The source function for $\lambda$\,4541\,{\AA} is
essentially Planckian in the relevant region.

\section{Discussion}
Quantitative spectroscopy has to avoid systematic errors in order to be
meaningful. A correct interpretation requires simultaneous reproduction of
all observational constraints (continua/spectral energy distribution,
line profiles). The line spectra of the most abundant elements,
hydrogen and helium, are particularly important in this.

We have investigated the suitability of hybrid non-LTE line-formation computations
for quantitative analyses of the hydrogen and helium line spectra of OB dwarf and giant stars. 
Our computations {\em simultaneously} reproduce the line spectra
throughout the visual and near-IR (where available) at high quality, as well as the
measured spectral energy distributions from the UV to near-IR, as discussed by
Nieva \& Przybilla~(\cite{np06a}). The only exceptions in our observational
sample are the cores of H$\alpha$ and
\ion{He}{ii} $\lambda$4686\,{\AA} in $\tau$\,Sco, because our computations
do not account for stellar winds. For two \ion{He}{i} lines blueward of
the traditionally analysed spectral region ($\sim$4000--5000\,{\AA}),
appropriate Stark broadening data is unavailable at present (see Table \ref{tablelines}).

Comparisons with other approaches, including the representative libraries of synthetic
spectra computed in non-LTE
(OSTAR2002 grid of Lanz \& Hubeny~\cite{la-hu03}) and LTE (Padova grid by
Munari et al.~\cite{mu03}) were also made.
We find good overall agreement between hybrid and full non-LTE calculations
within the parameter space investigated here: for the model atmosphere
structures, spectral energy distributions, and many detailed line profiles. 
Our hybrid approach is able to simultaneously reproduce the \ion{He}{i} singlets and triplets, 
confirmed by comparison with observations.
This is  not trivial and becomes important in particular when only few 
\ion{He}{i} lines are available in the observed
spectra. 

A comparison of TLUSTY$+$SYNSPEC with other state-of-the-art
non-LTE codes (FASTWIND and CMFGEN) for O-type stars by Najarro et al.~(\cite{Najarroetal06}) implies 
differences in all modelling results for the \ion{He}{i} singlets. 
However, while motivated by the problems with accurately reproducing
observations, this study does not present a detailed comparison of models and observed spectra\footnote {Our own tests indicate that both the ADS calculations and FASTWIND
provide high-quality fits to observation of the \ion{He}{i} singlet and
triplet spectra in the parameter space of interest here.}. Our work
indicates that the \ion{He}{i} singlet problem also persists 
in the hotter early B-type stars at solar metallicity 
(note that a significant reduction of metallicity alleviates the problem, as
Heap et al.~(\cite{Heapetal06}) find a simultaneous match of singlets and
triplets in SMC O stars).

Computations in LTE from the Padova grid, on the
other hand, systematically predict too shallow and/or too narrow line profiles.
In particular, the differences in the H$\gamma$ wings -- a common surface
gravity indicator -- result in systematically overestimated gravities by
up to $\sim$0.2\,dex in LTE (for fixed $T_\mathrm{eff}$). The differences in the equivalent
 widths of the H lines can
amount to up to $\sim$30\% and in the \ion{He}{i/ii} lines up
to a factor $>$2 compared to our non-LTE calculations, with the discrepancies
increasing with effective~temperature. Nevertheless we are not able to
quantify the differences in effective temperature determinations from
non-LTE and LTE ionization equilibria of \ion{He}{i/ii}, as some of the profiles
of the Padova grid do not reproduce observations even qualitatively.

 In terms of parameter range and the underlying physics, the hybrid non-LTE
approach is certainly restricted. It may be of limited use at higher temperatures 
(early and mid-O-type stars), lower gravities (early B-type and O-type supergiants), 
stars with strong winds, or extremely low metallicities.  
Nevertheless, the hybrid non-LTE approach is sufficient for studying normal OB dwarfs and
giants, as it allows the observed line spectra to be reproduced  
in the visual and near-IR over a wide range of atmospheric parameters.
Here it has advantages over other more sophisticated non-LTE techniques:
I) allows highly robust and detailed model atoms to be implemented and
to be tested efficiently (i.e. concentration on atomic data while avoiding
further complications like stellar winds),
e.g. for metals with hundreds of levels and thousands of transitions;
and II) the model calculations are fast: 
the computation of one H \& \ion{He}{i/ii} model with DETAIL$+$SURFACE takes only a few minutes  
on a modern PC (as of 2006).

\section{Conclusions}
A hybrid non-LTE approach, such as is often employed in the literature for
analyses of metal line spectra, has been \emph{thoroughly} tested for hydrogen and helium 
line formation in OB stars for the first time, using recently improved model atoms.
The synthetic spectra simultaneously match almost all measurable hydrogen and helium lines
in the optical and (where available) also in the near-IR spectral range of
six test stars covering a wide range of stellar parameters.

The comparison of state-of-the-art line-blanketed non-LTE and LTE models confirms that the 
atmospheric structure of OB dwarf and giant stars is described well under the
assumption of LTE, but not their spectral energy distribution and
also not their line spectra. 
For these stars in the range
20\,000\,K\,$\le$\,$T_\mathrm{eff}$\,$\le$\,35\,000\,K
and 3.0\,$\le$\,$\log g$\,$\le$\,4.5 (far from the Eddington limit), our hybrid non-LTE 
approach is equivalent to full hydrostatic non-LTE computations. 
 
It succeeds also in providing synthetic spectra that correctly reproduce the 
observed \ion{He}{i} singlet lines, avoiding inconsistencies recently reported
 in the literature.

In contrast to this, pure LTE modelling based on the Padova grid (or equivalent
computations with ATLAS9$+$SYNTHE) may give rise to considerable systematic
errors in the atmospheric parameter determination ($T_\mathrm{eff}$, $\log g$)
and to subsequent elemental abundance studies for the hotter stars in particular.
The problems with this approach may be remedied at least near the lower
temperature boundary by implementing proper line broadening data.

Finally, we suggest
that the \ion{He}{i} triplet be preferred for
analyses over the singlet lines from the OSTAR2002 grid
in order to avoid systematic uncertainties, in agreement with the findings of
 Najarro et al.~(\cite{Najarroetal06}). Hydrogen and helium 
line profiles from the Padova grid may be useful for
quantitative analyses of stars cooler than $\sim$22\,000\,K. Preference should be given
to the \ion{He}{i} $\lambda\lambda$3867, 4121, 4437, 4713, 5016, and
5048\,{\AA} transitions, which are least affected by non-LTE effects.

\begin{acknowledgements}

The authors wish to thank U.~Heber and K.~Cunha for their interest 
and their support of the project. We further thank K.~Butler for making 
DETAIL and SURFACE available; M.~Altmann, J. Puls, and P. A. Zaal for kindly providing the FEROS, 
the $K$-band, and the $\lambda$2.058\,$\mu$m data, respectively; and H. Edelmann for helping with the data reduction.
A special thanks go to U.H., K.B., K.C., J.P. and the anonymous referee of this paper for careful reading of the manuscript
and their comments and suggestions. M.F.N. acknowledges a DAAD scholarship. \\[1cm]
\end{acknowledgements}


\Online
\setcounter{table}{0}
\begin{table*}
\caption[]{Atomic data for H and \ion{He}{i/ii} line formation in the visual and NIR\\[-6mm]
\label{tablelines}}
\begin{tabular}{ll@{\hspace{1mm}--\hspace{1mm}}lrrlll}
\hline
$\lambda\,$(\AA) & \multicolumn{2}{c}{Transition} & $\log gf$ & Acc. & Src. &
Broad. & Comment\\[.8mm]
\hline
\ion{H}{i}:\\
3797.90 & 2                   & 10                  & $-$1.511 & AA & GRC  & SH\\
3835.38 & 2                   &  9                  & $-$1.362 & AA & GRC  & SH\\
3889.05 & 2                   &  8                  & $-$1.192 & AA & GRC  & SH\\
3970.07 & 2                   &  7                  & $-$0.993 & AA & GRC  & SH & blend with interstellar \ion{Ca}{ii} H\\
4101.73 & 2                   &  6                  & $-$0.753 & AA & GRC  & SH\\
4340.46 & 2                   &  5                  & $-$0.447 & AA & GRC  & SH\\
4861.32 & 2                   &  4                  & $-$0.020 & AA & GRC  & SH\\
6562.80 & 2                   &  3                  &    0.710 & AA & GRC  & SH\\
8413.32 & 3                   & 19                  & $-$1.823 & AA & GRC  & SH\\
8437.96 & 3                   & 18                  & $-$1.748 & AA & GRC  & SH\\
8467.26 & 3                   & 17                  & $-$1.670 & AA & GRC  & SH\\
8502.49 & 3                   & 16                  & $-$1.586 & AA & GRC  & SH\\
8545.39 & 3                   & 15                  & $-$1.495 & AA & GRC  & SH\\
8598.39 & 3                   & 14                  & $-$1.398 & AA & GRC  & SH\\
8665.02 & 3                   & 13                  & $-$1.292 & AA & GRC  & SH\\
8750.47 & 3                   & 12                  & $-$1.175 & AA & GRC  & SH\\[1mm]
\ion{He}{ii}:\\
3796.34 & 4                   & 20                  & $-$1.487 & AA & GRC  & G60, G67 & blend with H$\vartheta$\\
3813.50 & 4                   & 19                  & $-$1.414 & AA & GRC  & G60, G67 & \\
3833.81 & 4                   & 18                  & $-$1.337 & AA & GRC  & G60, G67 & blend with H$\eta$\\
3858.08 & 4                   & 17                  & $-$1.255 & AA & GRC  & G60, G67 & \\
3887.45 & 4                   & 16                  & $-$1.166 & AA & GRC  & G60, G67 & blend with H$\zeta$\\
3923.49 & 4                   & 15                  & $-$1.071 & AA & GRC  & SB$^{\rm a}$ & blend with \ion{He}{i}\\
3968.44 & 4                   & 14                  & $-$0.967 & AA & GRC  & SB$^{\rm a}$ & blend with H$\epsilon$\\
4025.61 & 4                   & 13                  & $-$0.852 & AA & GRC  & SB$^{\rm a}$ & blend with \ion{He}{i}\\
4100.05 & 4                   & 12                  & $-$0.725 & AA & GRC  & SB & blend with H$\delta$\\
4199.84 & 4                   & 11                  & $-$0.582 & AA & GRC  & SB & blend with \ion{N}{iii}\\
4338.67 & 4                   & 10                  & $-$0.417 & AA & GRC  & SB & blend with H$\gamma$\\
4541.59 & 4                   &  9                  & $-$0.223 & AA & GRC  & SB\\
4685.70 & 3                   &  4                  &    1.181 & AA & GRC  & SB\\
4859.32 & 4                   &  8                  &    0.014 & AA & GRC  & SB & blend with H$\beta$\\
5411.52 & 4                   &  7                  &    0.321 & AA & GRC  & SB\\
6560.09 & 4                   &  6                  &    0.759 & AA & GRC  & SB & blend with H$\alpha$\\[1mm]
\hline
\end{tabular}\\
$^{\rm a}$ unpublished, priv. comm.;~~~$^{\rm b}$ vacuum wavelengths\\[1mm]
accuracy indicators -- uncertainties within:~~AA: 1\%; A: 3\%\\[1mm]
sources of $gf$-values:~~
FTS: Fernley et al.~(\cite{Fernleyetal87});~
GRC: Green et al.~(\cite{Greenetal57});~
SPA: Schiff et al.~(\cite{Schiffetal71})\\[1mm]
sources for Stark broadening parameters:~~
BCS69: Barnard et al.~(\cite{Barnardetal69});~
C:     Cowley~(\cite{Cowley71});~
DSB:   Dimitri\-jevi\'c~\& Sahal-Br\'echot~(\cite{DiSa90});~
G60:   Griem~(\cite{Griem60});~
G64:   Griem~(\cite{Griem64});~
G67:   Griem~(\cite{Griem67});~
GBKO:  Griem et al.~(\cite{Griemetal62});~
S:     Shamey~(\cite{Shamey69});~
SB:    Sch\"oning \& Butler~(\cite{SchBu89});~
SH:    Stehl\'e \& Hutcheon~(\cite{sh99})
\end{table*}
\newpage
\begin{table*}
\noindent{\bf Table~\ref{tablelines}.} continued\\[1mm]
\begin{tabular}{ll@{\hspace{1mm}--\hspace{1mm}}lrrlll}
\hline
$\lambda\,$(\AA) & \multicolumn{2}{c}{Transition} & $\log gf$ & Acc. & Src. &
Broad. & Comment\\[.8mm]
\hline
\ion{He}{i}:\\
3819.60 & 2p\,$^3$P$^{\rm o}$ & 6d\,$^3$D           & $-$0.931 &  A & FTS  & DSB & broadening data to be improved, forbidden components missing\\
3819.61 & 2p\,$^3$P$^{\rm o}$ & 6d\,$^3$D           & $-$1.153 &  A & FTS  & DSB &{\ldots} \\
3819.76 & 2p\,$^3$P$^{\rm o}$ & 6d\,$^3$D           & $-$1.630 &  A & FTS  & DSB &{\ldots} \\
3867.47 & 2p\,$^3$P$^{\rm o}$ & 6s\,$^3$S           & $-$2.037 &  A & FTS  & DSB\\
3867.48 & 2p\,$^3$P$^{\rm o}$ & 6s\,$^3$S           & $-$2.260 &  A & FTS  & DSB\\
3867.63 & 2p\,$^3$P$^{\rm o}$ & 6s\,$^3$S           & $-$2.737 &  A & FTS  & DSB\\
3888.60 & 2s\,$^3$S           & 3p\,$^3$P$^{\rm o}$ & $-$1.668 & AA & SPA  & G64   & near core of H$\zeta$\\
3888.65 & 2s\,$^3$S           & 3p\,$^3$P$^{\rm o}$ & $-$0.765 & AA & SPA  & G64   &{\ldots}\\
3926.54 & 2p\,$^1$P$^{\rm o}$ & 8d\,$^1$D           & $-$1.652 &  A & FTS  & DSB & broadening data to be improved, blends by \ion{Si}{iii} \& \ion{S}{ii/iii} \\
3935.95 & 2p\,$^1$P$^{\rm o}$ & 8s\,$^1$S           & $-$2.772 &  A & FTS  & DSB\\
3964.73 & 2s\,$^1$S           & 4p\,$^1$P$^{\rm o}$ & $-$1.290 &  A & FTS  & G64   & in wing of H$\epsilon$\\
4009.26 & 2p\,$^1$P$^{\rm o}$ & 7d\,$^1$D           & $-$1.449 &  A & FTS  & DSB\\
4023.98 & 2p\,$^1$P$^{\rm o}$ & 7s\,$^1$S           & $-$2.572 &  A & FTS  & DSB\\
4026.18 & 2p\,$^3$P$^{\rm o}$ & 5d\,$^3$D           & $-$2.600 &  A & FTS  & S\\
4026.19 & 2p\,$^3$P$^{\rm o}$ & 5d\,$^3$D           & $-$0.633 &  A & FTS  & S\\
4026.20 & 2p\,$^3$P$^{\rm o}$ & 5d\,$^3$D           & $-$0.851 &  A & FTS  & S\\
4026.36 & 2p\,$^3$P$^{\rm o}$ & 5d\,$^3$D           & $-$1.328 &  A & FTS  & S\\
4120.81 & 2p\,$^3$P$^{\rm o}$ & 5s\,$^3$S           & $-$1.722 &  A & FTS  & GBKO & blends with \ion{O}{ii}, \ion{C}{iii}, \& \ion{Fe}{iii}\\
4120.82 & 2p\,$^3$P$^{\rm o}$ & 5s\,$^3$S           & $-$1.945 &  A & FTS  & GBKO &{\ldots}\\
4120.99 & 2p\,$^3$P$^{\rm o}$ & 5s\,$^3$S           & $-$2.422 &  A & FTS  & GBKO &{\ldots} \\
4143.76 & 2p\,$^1$P$^{\rm o}$ & 6d\,$^1$D           & $-$1.203 &  A & FTS  & DSB & numerous blends with \ion{O}{ii} \& \ion{N}{ii}\\
4168.97 & 2p\,$^1$P$^{\rm o}$ & 6s\,$^1$S           & $-$2.332 &  A & FTS  & DSB & strong blend with \ion{O}{ii}\\
4387.93 & 2p\,$^1$P$^{\rm o}$ & 5d\,$^1$D           & $-$0.886 &  A & FTS  & S\\
4437.55 & 2p\,$^1$P$^{\rm o}$ & 5s\,$^1$S           & $-$2.026 &  A & FTS  & GBKO & continuum may be affected by diffuse interstellar band\\
4471.47 & 2p\,$^3$P$^{\rm o}$ & 4d\,$^3$D           & $-$0.210 &  A & FTS  & BCS69\\
4471.49 & 2p\,$^3$P$^{\rm o}$ & 4d\,$^3$D           & $-$0.432 &  A & FTS  & BCS69\\
4471.68 & 2p\,$^3$P$^{\rm o}$ & 4d\,$^3$D           & $-$0.909 &  A & FTS  & BCS69\\
4713.14 & 2p\,$^3$P$^{\rm o}$ & 4s\,$^3$S           & $-$1.276 &  A & FTS  & G64\\
4713.16 & 2p\,$^3$P$^{\rm o}$ & 4s\,$^3$S           & $-$1.498 &  A & FTS  & G64\\
4713.38 & 2p\,$^3$P$^{\rm o}$ & 4s\,$^3$S           & $-$1.976 &  A & FTS  & G64\\
4921.93 & 2p\,$^1$P$^{\rm o}$ & 4d\,$^1$D           & $-$0.442 &  A & FTS  & BCS69 & treatment of forbidden component to be improved\\
5015.68 & 2s\,$^1$S           & 3p\,$^1$P$^{\rm o}$ & $-$0.820 & AA & SPA  & G64\\
5047.74 & 2p\,$^1$P$^{\rm o}$ & 4s\,$^1$S           & $-$1.600 &  A & FTS  & GBKO\\
5875.60 & 2p\,$^3$P$^{\rm o}$ & 3d\,$^3$D           & $-$1.511 &  A & FTS  & G64\\
5875.61 & 2p\,$^3$P$^{\rm o}$ & 3d\,$^3$D           &    0.480 &  A & FTS  & G64\\
5875.63 & 2p\,$^3$P$^{\rm o}$ & 3d\,$^3$D           & $-$0.338 &  A & FTS  & G64\\
5875.64 & 2p\,$^3$P$^{\rm o}$ & 3d\,$^3$D           &    0.138 &  A & FTS  & G64\\
5875.97 & 2p\,$^3$P$^{\rm o}$ & 3d\,$^3$D           & $-$0.214 &  A & FTS  & G64\\
6678.15 & 2p\,$^1$P$^{\rm o}$ & 3d\,$^1$D           &    0.328 &  A & FTS  & G64\\
7065.18 & 2p\,$^3$P$^{\rm o}$ & 3s\,$^3$S           & $-$0.458 &  A & FTS  & GBKO & weak telluric line contamination\\
7065.22 & 2p\,$^3$P$^{\rm o}$ & 3s\,$^3$S           & $-$0.680 &  A & FTS  & GBKO &{\ldots}\\
7065.71 & 2p\,$^3$P$^{\rm o}$ & 3s\,$^3$S           & $-$1.157 &  A & FTS  & GBKO &{\ldots}\\
7281.35 & 2p\,$^1$P$^{\rm o}$ & 3s\,$^1$S           & $-$0.854 &  A & FTS  & GBKO & strong telluric line contamination\\
10829.09 & 2s\,$^3$S & 2p\,$^3$P$^{\rm o}$          & $-$0.745 & AA & SPA  & G64\\
10830.25 & 2s\,$^3$S & 2p\,$^3$P$^{\rm o}$          & $-$0.268 & AA & SPA  & G64\\
10830.34 & 2s\,$^3$S & 2p\,$^3$P$^{\rm o}$          & $-$0.046 & AA & SPA  & G64\\
20586.92$^{\rm b}$ & 2s\,$^1$S & 2p\,$^1$P$^{\rm o}$ & $-$0.424 & AA & SPA & DSB & strong telluric line contamination\\
21125.79$^{\rm b}$ & 3p\,$^3$P$^{\rm o}$ & 4s\,$^3$S & $-$0.138 &  A & FTS & DSB\\
21125.89$^{\rm b}$ & 3p\,$^3$P$^{\rm o}$ & 4s\,$^3$S & $-$0.360 &  A & FTS & DSB\\
21127.09$^{\rm b}$ & 3p\,$^3$P$^{\rm o}$ & 4s\,$^3$S & $-$0.837 &  A & FTS & DSB\\
21137.80$^{\rm b}$ & 3p\,$^1$P$^{\rm o}$ & 4s\,$^1$S & $-$0.527 &  A & FTS & DSB\\[1mm]
\hline
\end{tabular}\\
\end{table*}

\clearpage

\begin{figure*}
\centering
\includegraphics[width=0.97\linewidth]{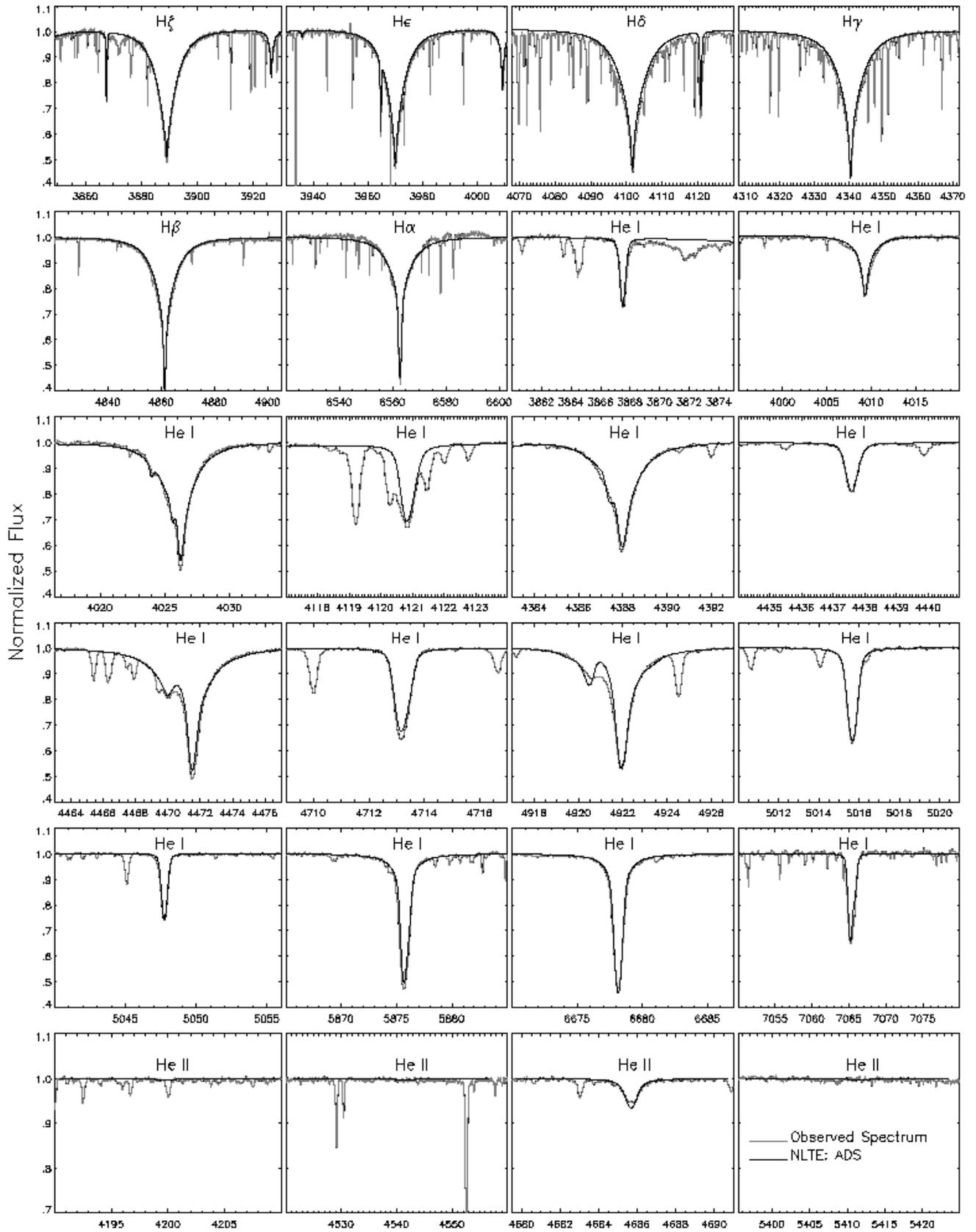}\\[-6mm]
\caption{Line fits for HR\,1861 (B1\,IV). For atmospheric parameters see
Table~\ref{parameters}, and for further discussion see the text.}
\label{visualHR1861}
\end{figure*}

\clearpage

\begin{figure*}
\centering
\includegraphics[width=0.97\linewidth]{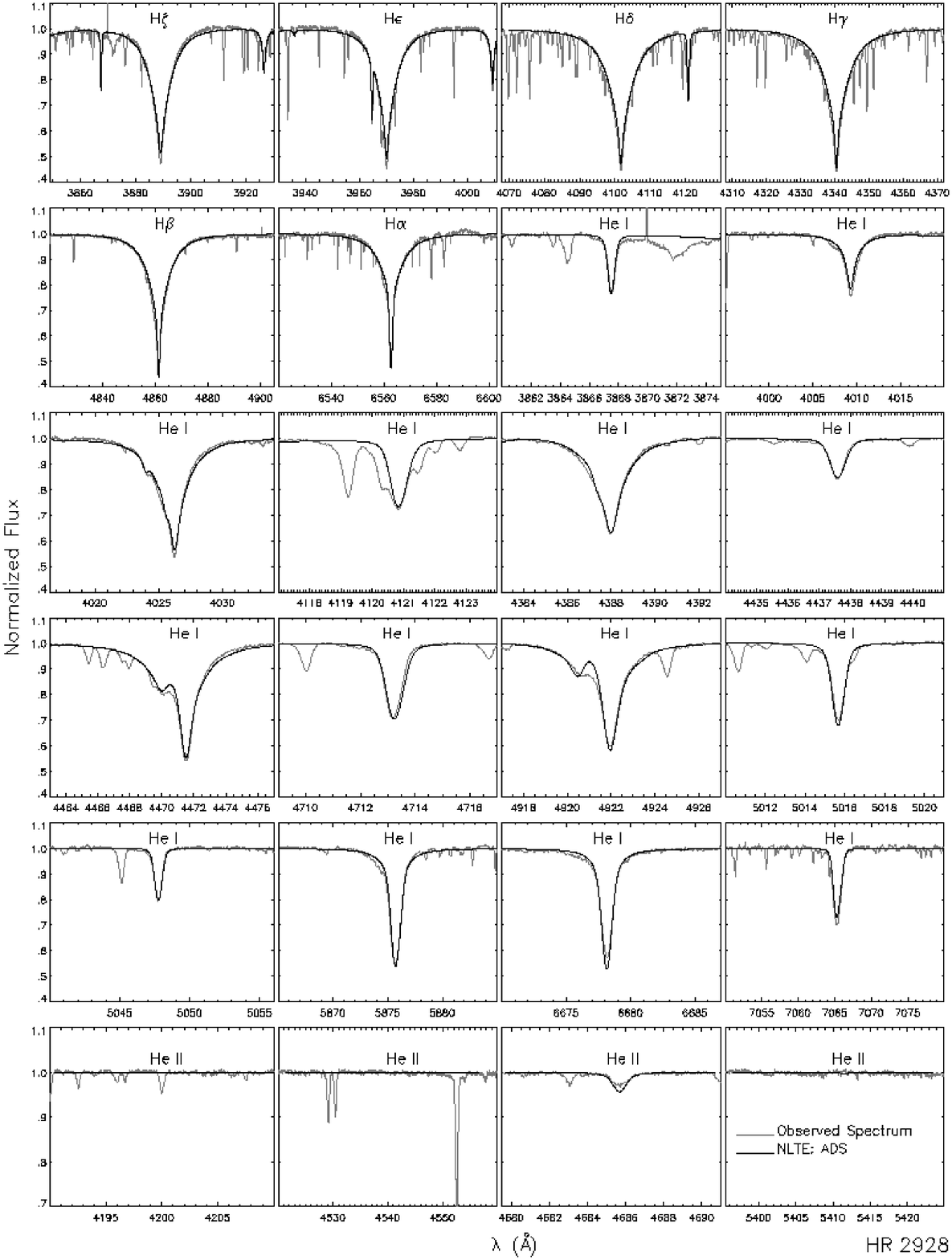}\\[-6mm]
\caption{As Fig.~\ref{visualHR1861}, but for HR\,2928 (B1\,IV).}
\label{visualHR2928}
\end{figure*}

\clearpage

\begin{figure*}
\centering
\includegraphics[width=0.97\linewidth]{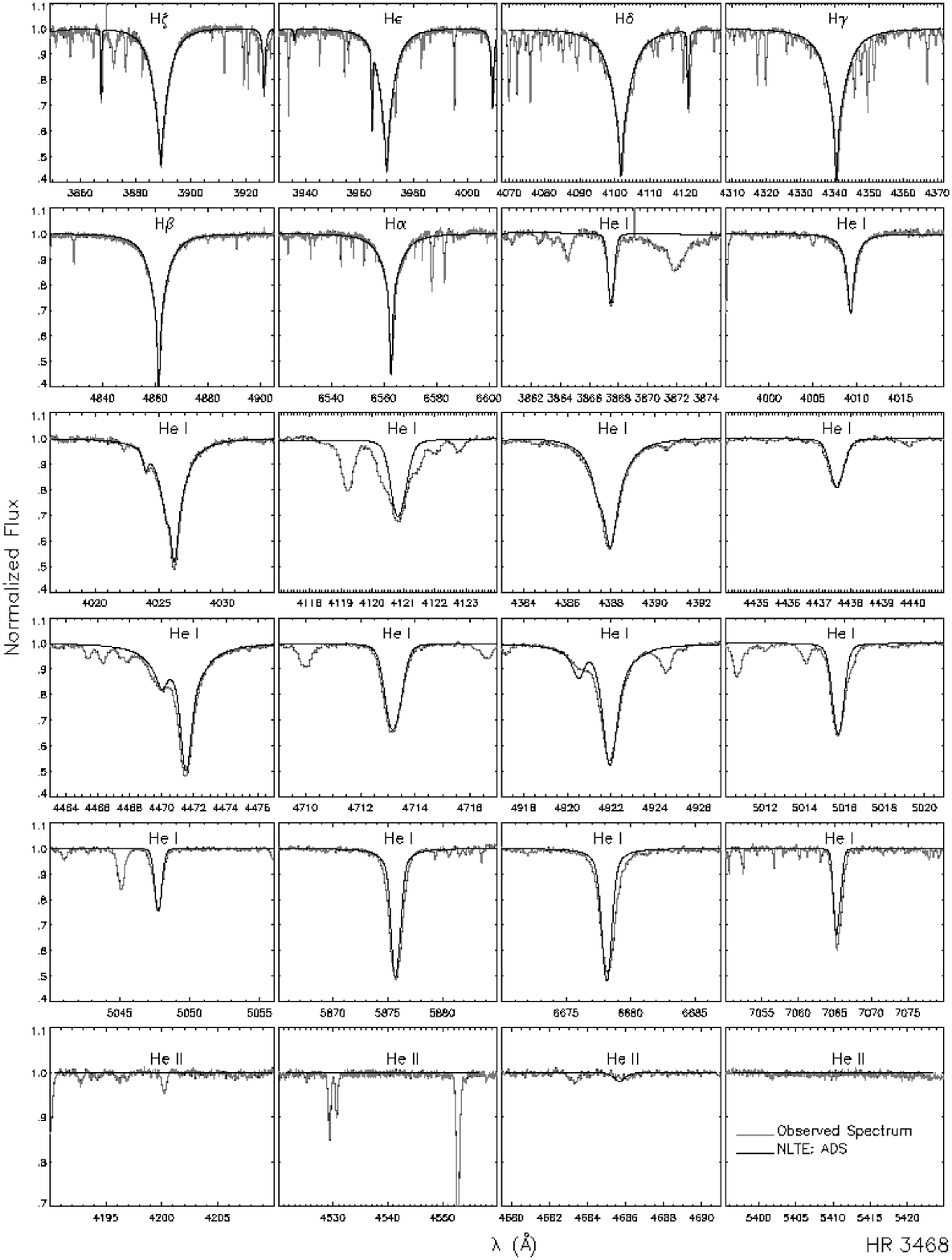}\\[-6mm]
\caption{As Fig.~\ref{visualHR1861}, but for HR\,3468 (B1.5\,III).}
\label{visualHR3468}
\end{figure*}

\clearpage

\begin{figure*}
\centering
\includegraphics[width=0.97\linewidth]{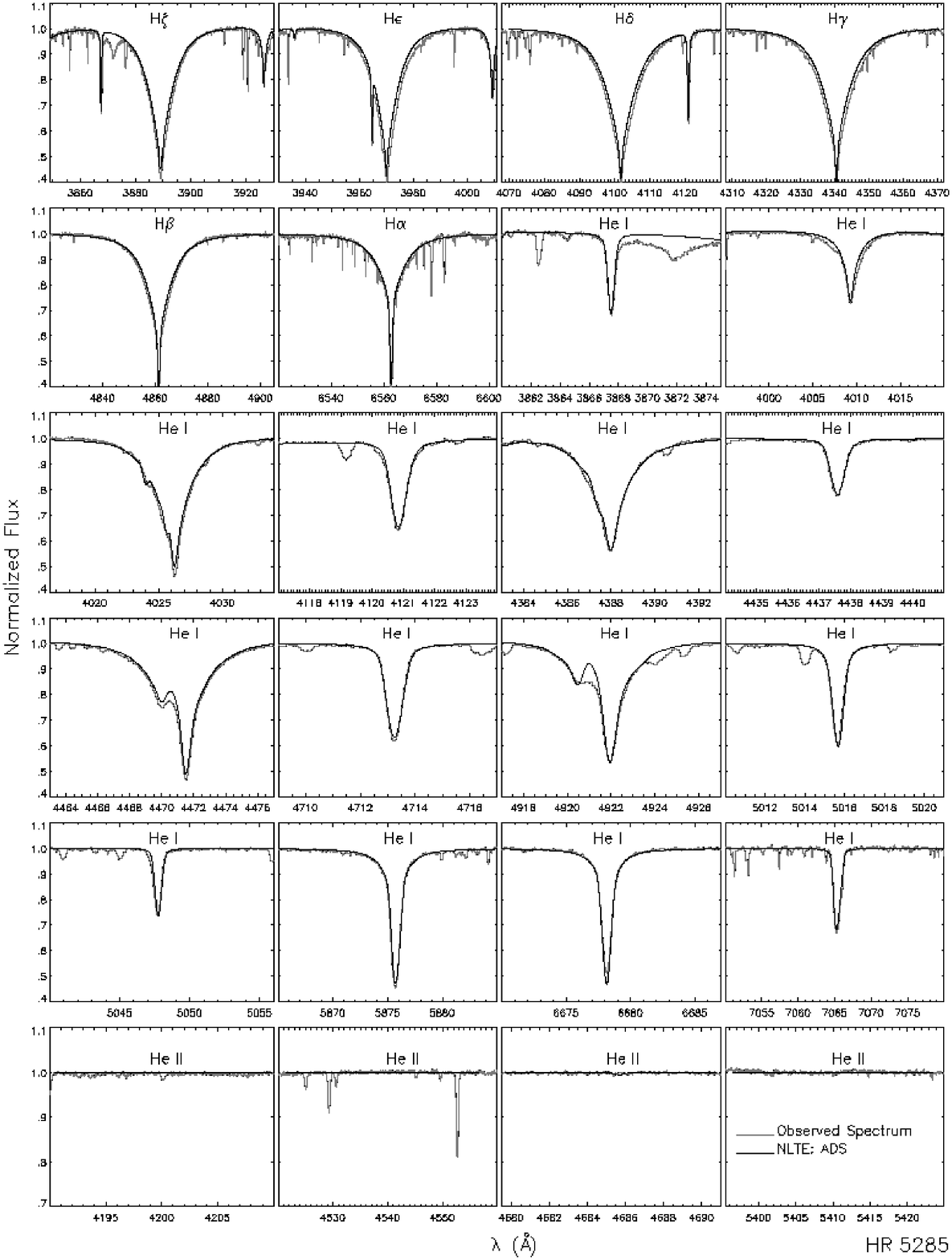}\\[-6mm]
\caption{As Fig.~\ref{visualHR1861}, but for HR\,5285 (B2\,V). }
\label{visualHR5285}
\end{figure*}

\clearpage

\end{document}